\documentclass[10pt]{article}

\usepackage{amsmath}

\usepackage{array}

\usepackage{appendix}

\usepackage{tocloft}                   

\usepackage{graphicx}

\usepackage{amsfonts}

\usepackage{amssymb}

\usepackage{mathrsfs}

\usepackage{yfonts}

\usepackage{euscript}

\usepackage{centernot}                 

\usepackage{ifsym}                     

\usepackage{upgreek}

\usepackage{mathtools}

\usepackage{color}

\usepackage{slantsc}
\usepackage{calligra}

\usepackage{bbold}          

\usepackage[T1]{fontenc}

\usepackage{epsf}

\usepackage{latexsym}

\usepackage{tipa}

\usepackage{makeidx}

\makeindex

\def\b{\begin{equation}}

\def\e{\begin{equation}}

\def\be{\begin{equation}}              

\def\ee{\end{equation}}

\def\beq{\begin{equation}}

\def\eeq{\end{equation}}

\def\bea{\begin{eqnarray}}

\def\eea{\end{eqnarray}}

\def\m{\mbox{ }}

\def\mma {\m , \m \m }

\def\!{\hspace{-1.6667em}}

\def\n{\noindent}

\def\u{\underline}

\def\w{\widetilde}

\def\s{\stackrel}

\def\biN{\mbox{\boldmath$N$}}

\def\disjoint{\mbox{\scriptsize $\coprod$}}

\def\mA{\mbox{A}}  

\def\mB{\mbox{B}}  

\def\mC{\mbox{C}}                        

\def\mD{\mbox{D}}                        

\def\mE{\mbox{E}}                        

\def\mF{\mbox{F}}

\def\mG{\mbox{G}}

\def\mH{\mbox{H}} 

\def\mK{\mbox{K}}

\def\mP{\mbox{P}}

\def\mQ{\mbox{Q}}

\def\mS{\mbox{S}}                        

\def\mT{\mbox{T}} 

\def\mW{\mbox{W}}

\def\mb{\mbox{b}}

\def\md{\mbox{d}} 

\def\me{\mbox{e}}

\def\ml{\mbox{l}}

\def\mo{\mbox{o}}

\def\mp{\mbox{p}}

\def\ms{\mbox{s}}

\def\bupSigma{\mbox{\boldmath$\Sigma$}}                 

\def\sa{\mbox{\scriptsize a}}

\def\sm{\mbox{\scriptsize m}}

\def\sx{\mbox{\scriptsize x}}

\def\sB{\mbox{\scriptsize B}}

\def\sD{\mbox{\scriptsize D}}

\def\sG{\mbox{\scriptsize G}}

\def\sO{\mbox{\scriptsize O}}

\def\sP{\mbox{\scriptsize P}}

\def\sumi2{\sum\mbox{}_{\mbox{}_{\mbox{\scriptsize $i$=1}}}^2}

\def\sumi3{\sum\mbox{}_{\mbox{}_{\mbox{\scriptsize $i$=1}}}^3}

\def\sumABcycles3{\sum\mbox{}_{\mbox{}_{\mbox{\scriptsize cycles $A,B$=1}}}^{3}}

\def\sumCDcycles3{\sum\mbox{}_{\mbox{}_{\mbox{\scriptsize cycles $C,D$=1}}}^{3}}

\def\sumj3{\sum\mbox{}_{\mbox{}_{\mbox{\scriptsize $j$=1}}}^3}

\def\sumk3{\sum\mbox{}_{\mbox{}_{\mbox{\scriptsize $k$=1}}}^3}

\def\prodiA1{\prod\mbox{}_{\mbox{}_{\mbox{\scriptsize $i$=1}}}^{A - 1}}

\def\bigtimes{\mbox{\Large $\times$}}

\def\d{\textrm{d}}                                                  

\def\es{\m = \m}

\def\:={\m := \m}

\def\=:{\m =: \m}

\def\FrA{\mbox{$\mathfrak{A}$}}                                

\def\FrT{\mathfrak{T}}                                         

\def\FrC{\mbox{$\mathfrak{C}$}}                                
                                                       
\def\FrX{\mathfrak{X}}                                         
\def\sFrR{\mbox{\scriptsize $\mathfrak{R}$}}                   
\def\FrS{\mbox{\Large $\mathfrak{s}$}}                         
\def\sFrS{\mbox{\large$\mathfrak{s}$}}                         

\def\sFrM{\mbox{\scriptsize$\mathfrak{M}$}}                    

\def\lFrg{\mbox{\Large$\mathfrak{g}$}}                         
\def\FrH{\mbox{$\mathfrak{H}$}}                                
\def\FrT{\mbox{\boldmath$\mathfrak{T}$}}                       
\def\sFrT{\mbox{\scriptsize$\mathfrak{T}$}}                    

\def\Hilb{\mbox{{\boldmath$\mathfrak{H}$}ilb}}                 
\def\bFrB{\mbox{\boldmath$\mathfrak{B}$}}                      
                                                               
\def\Big{\bFrB\mbox{ig}}

\def\Grand{\bFrG\mbox{rand}}

\def\FrQ{\mbox{\Large $\mathfrak{q}$}}                               
\def\bFrC{\mbox{\boldmath$\mathfrak{C}$}}                            
\def\bFrL{\mbox{\boldmath$\mathfrak{L}$}}                            

\def\Phase{\mbox{{\boldmath$\mathfrak{P}$}hase}}                     

\def\bFrR{\mbox{\boldmath$\mathfrak{R}$}}                            
\def\Rig-Phase{\bFrR\mbox{ig-}\Phase}                                
                                                                													   
\def\FrR{\mbox{\boldmath$\mathfrak{R}$}}                             

\def\sFrR{\mbox{\scriptsize\boldmath$\mathfrak{R}$}}                 
\def\bFrR{\mbox{\boldmath$\mathfrak{R}$}}                            

\def\bFrR{\mbox{\boldmath$\mathfrak{R}$}}                            

\def\1mat{\u{\u{1}}}                                                 

\def\Leib{\bFrL\mbox{eib}}                                           

\def\Co{\bFrC\mo}                                                    

\def\Res{\bFrR\me\ms}                                                

\def\Positive-Modespace{\mbox{{\boldmath$\mathfrak{M}$}odespace$^+$}}

\def\POSITIVE-MODESPACE{\mbox{{\boldmath$\mathfrak{M}$}ODESPACE$^+$}}
                                                                                                                             														
\def\bFrS{\mbox{\Large $\mathfrak{s}$}}                              
			
\def\bFrG{\mbox{ $\mathfrak{G}$}}                                    %
\def\Riem{\bFrR\mbox{iem}}                                           
\def\Superspace{\bFrS\mbox{uperspace}}                               

\def\FrO{\mbox{$\mathfrak{O}$}}                                      

\def\Top{\FrT\mo\mp}

\def\Rel{\FrR\me\ml}

\def\lattice{\mbox{\bf\Large$\mathfrak{L}$}}                                      

\def\sFrC{\mbox{\boldmath\scriptsize$\mathfrak{C}$}}                        

\def\Kin-Hilb{\mbox{{\boldmath$\mathfrak{K}$}in-\Hilb}}                     

\def\Mid-Hilb{\mbox{{\boldmath$\mathfrak{M}$}id-\Hilb}}                     

\def\Dyn-Hilb{\mbox{{\boldmath$\mathfrak{D}$}yn-\Hilb}}                     

\def\5Star{\mbox{\Large$\star$}}              

\begin{document}

\begin{titlepage}

\begin{center}

\Large{\bf Rubber Relationalism:} \normalsize

\vspace{0.1in}

\large{\bf Smallest Graph-Theoretically Nontrivial Leibniz Spaces}

\vspace{0.2in}

{\normalsize \bf Edward Anderson$^*$}

\vspace{.2in}

\end{center}

\begin{abstract}

\n Kendall's Similarity Shape Theory for constellations of N points in the carrier space $\mathbb{R}^d$ as quotiented by the similarity group 
was developed for use in Probability and Statistics. 
It was subsequently shown to reside within (Classical and Quantum) Mechanics' Shape-and-Scale Theory, 
in which points are interpreted as particles, carrier space plays the role of absolute space, and the Euclidean group is quotiented out.  
Let us jointly refer to Shape Theory and Shape-and-Scale Theory as Relational Theory, 
and to the corresponding reduced configuration spaces as relational spaces.
We now consider a less structured version: the Topological Relational Theory of `rubber configurations'. 
This already encodes some features of the much more diverse Geometrical Relational Theories; 
in contrast with the latter's (stratified) manifold relational spaces, the former's are graphs: much simpler to visualize and study.  
These graphs' edges encode the `topological adjacency' relation.  
We concentrate on Leibniz spaces, corresponding to indistinguishable points and mirror-image identification. 
These are moreover the building blocks of the distinguishable and (where possible) mirror-image distinct cases' relational spaces.  
For connected manifold without boundary models of carrier space, there are just 3 `rubber relationalisms': 
$\mathbb{R}$, $\mathbb{S}^1$, and a joint one for all carrier spaces with $d \geq 2$.   
For $d \geq 2$, rubber configurations are in 1 : 1 correspondence with partitions, with the $\mathbb{S}^1$ and $\mathbb{R}$ cases giving successive refinements.
We find that generic and maximal configurations are universally present as cone points, as are binaries in the first 2 cases. 
Deconing leaves us with residue graphs containing the $N$-specific information. 
We provide graph-theoretical nontriviality criteria for which $N = 6$, 6 and 5 are minimal across these models, 
and stronger such for which $N = 8$, 8 and 6 are minimal, 
and outline GR topology-change analogue-model and $N$-body problem applications.  

\end{abstract}

\n {\bf PACS}: 04.20.Cv, 02.40.Pc.

\m 

\n {\bf Physics keywords}: $N$-Body Problem, Background Independence, Topological Background Independence, configuration spaces. 

\m

\n {\bf Mathematics keywords}: Shape Theory, Applied Graph Theory, Applied Topology, Shape Statistics.
 
\vspace{0.1in}
    
\n $^*$ Dr.E.Anderson.Maths.Physics@protonmail.com

\vspace{0.1in}

{            \begin{figure}[!ht]
\centering
\includegraphics[width=0.8\textwidth]{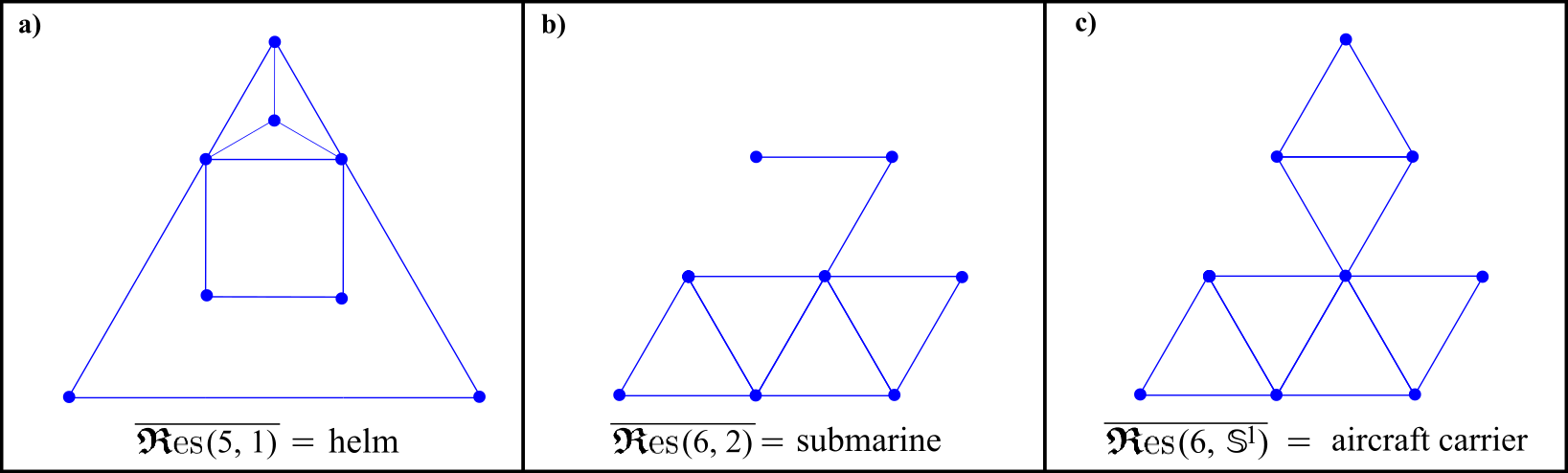}
\caption[Text der im Bilderverzeichnis auftaucht]{        \footnotesize{The first nontrivial residue graph complements for the three universality classes of rubber Leibniz spaces 
are the $N = 5$ `helm' in $\mathbb{R}$, 
    the $N = 6$ `submarine' in $\geq 2$,  
and the $N = 6$ `aircraft carrier in $\mathbb{S}^1$. 
} }
\label{Top-2-Front} \end{figure}          }

\end{titlepage}

\section{Introduction}

{\bf Motivation} 

\m 

\n Shape Theory in David Kendall's sense \cite{Kendall84, Kendall89, Kendall} models space as $\mathbb{R}^d$ of dimension $d$, 
and treats constellations of $N$ points thereupon by quotienting out the similarity group of transformations, $Sim(d)$.   
This involves a metric notion of shape, and concentrates on the configuration space formed by these metric shapes: {\it shape space}, 
\be 
\FrS(d, \, N)   \:=  \frac{\bigtimes_{i = 1}^N \mathbb{R}^d}{Sim(d)}  
             \es  \frac{\mathbb{R}^{N \, d}}{Sim(d)}                  \m .  
\label{Sim-S}			 
\ee 
This work is most familiar in the Shape Statistics literature \cite{Small, Kendall, Bhatta, DM16, PE16}, 
in which probability measures and statistics are set up in concordance with the shape space's geometry.  

\m 

\n See moreover e.g.\ \cite{Smale70, LR95, LR97, M02-M05, RPM, AF, FileR, APoT, ABook, I, II, III, III-Concat} for related work in other fields, 
including Mechanics, Quantization and modelling some aspects of Classical and Quantum General Relativity's Background Independence \cite{A64-A67, I93, Giu06, ABook} 
and Problem of Time \cite{K92, I93, APoT, ABook}.  
Some of these further works consider quotienting instead by the Euclidean group of transformations, $Eucl(d)$. 
This gives Shape-and-Scale Theory. 
The corresponding configuration space for this is {\it shape-and-scale space}, 
\be 
\FrR(d, \, N)  \es  \frac{\mathbb{R}^{N \, d}}{Eucl(d)} 
                = \mC(\FrS(d, \, N))                                      \m .   
\label{Eucl-R}				 
\ee
\n Shape space is furtherly significant as a subspace of shape-and-scale space, 
and moreover one which is geometrically simpler and a useful intermediary to construct first, 
with shape-and-scale space then being the cone over shape space \cite{LR97, Cones, FileR}.
This is the meaning of the last equality in (\ref{Eucl-R}).
We collectively refer to Shape(-and-Scale) Theory as Relational Theory and to the corresponding configuration spaces -- shape(-and-scale) spaces -- as relational spaces.
`Relational' is meant here in the sense of the Absolute versus Relational Debate \cite{Newton, L, M, BB82-DoD-Buckets, ABook} which dates at least as far back as Newton versus Leibniz.   

\m

\n On the one hand, in physical applications, the points are often considered to be particles, so we use `point-or-particle' as a portmanteau name and concept.
On the other hand, in the case of statistical applications, the points represent location data.  

\m 

\n Three further ambiguities in the above modelling, which substantially further enrich both its foundational scope and its applicability, are as follows.
Together, these create a sizeable field of study: generalized Kendall-type Geometrical Shape(-and-Scale) Theory.  

\m 

\n 1) $\mathbb{R}^d$'s role of absolute space, or, more generally of a carrier space (i.e.\ a not necessarily physically realized counterpart)  
can be allotted to other models of space instead, such as $\mathbb{S}^d$, $\mathbb{T}^d$ or $\mathbb{RP}^d$.    

\m 

\n 2) The role of the continuous group being quotiented out is more generally that of a group of automorphisms which are held to be irrelevant to the modelling in question.  
This implements part of what is termed Background Independence in Theoretical Physics and the Foundations of Physics \cite{A64-A67, I93, Giu06, ABook}. 
Alternatives to $Sim(d)$ and $Eucl(d)$ here include  
the affine group \cite{Sparr-GT09, Bhatta, PE16},  
the conformal group \cite{AMech, ABook}.  
and the projective general linear \cite{MP05-KKH16, Bhatta, PE16}. 

\m 

\n 3) Discrete transformations can furthermore be quotiented out so as to incorporate one or both of mirror-image and label indistinguishabilities 
\cite{I, II, III} into the modelling.   
Leibniz spaces is the most quotieneted-out configuration spaces; here just 1 copy of each unlabelled (scaled) shape is present.  
Leibniz spaces are moreover further motivated as the building blocks 
of the distinguishable and (where possible) mirror image distinct cases' (scaled) shape spaces \cite{Top-Shapes, ACirc}.   
In the geometrical case, these larger (scaled) shape spaces are tessellations whose individual tiles are Leibniz spaces \cite{Kendall89, AF, FileR, II, III}, 
whereas \cite{Top-Shapes} exhibits a similar pattern repetition withing the rubber case's graphs. 
In the current article, we thus cut down on graph orders by considering the topological Leibniz space building blocks themselves, 
and asking how -- and for which minimal $N$ -- these themselves become nontrivial. 
The nontriviality of these graphs then pervades all the further topological and geometrical graphs underlied by these Leibniz space building blocks 
(the topological Leibniz spaces encoding part of the structure of their geometrical counterparts).  

\vspace{10in}

\n{\bf Outline of the rest of this article} 

\m 

\n A more detailed account of 1) and 3) is given in Secs 2 and 3.    
The current Article moreover concentrates rather on a coarser view universal within the above pletora of Relational Theories. 
I.e.\ Sec \ref{Rubber}'s Topological Relational Theory of rubber (scaled) shapes.  
This maintains significant distinction between the $d = 1$ and $d \geq 1$ versions, but has no further dependence on spatial dimension.  
It is also independent of $\lFrg$ modulo whether or not this includes scale [thus 2) does not enter the current paper further, 
other than through the current paper's results undelying some aspects of Relational Theory for {\sl whichever} $\lFrg$].  
In Sec \ref{Partitions}, we recollect the $(d, \, N)$ = (1, 3), (1, 4), (2, 3), (2, 4) cases from \cite{I, II, III, Top-Shapes}.  

\m 

\n For $d \geq 2$, the rubber scaled shapes are in 1 : 1 correspondence with the partitions (Sec \ref{Partitions}). 
To Geometric Relational Theories' reduced configuration spaces being stratified manifolds in general, Topological Relational Theories' are just graphs.
The edges in question encode topological adjacency, and are in general present in excess of the lattice of partition refinements' edges (Sec \ref{Not-Enough}).  
These results are furthermore independent of choice of carrier space for $d \geq 2$. 

\m

\n On the other hand, in 1-$d$ in a fine-graining of the partitions occurs for sufficiently large $N$; 
this is moreover finer-grained for $\FrC^d = \mathbb{R}$ than for $\mathbb{S}^1$.  
The first $\mathbb{R}$ to $\FrC^{\geq 2}$ distinction is for $N = 4$ (Sec \ref{5,1}) and the first $\mathbb{S}^1$ distinction occurs for $N = 6$ (Sec \ref{6,S}).  
All in all, there are 3 rubber classes for connected manifold without boundary carrier spaces: $\mathbb{R}$, $\mathbb{S}^1$ and $\FrC^d$ for $d \geq 2$. 
As well as being much more general and much more relevant, $d \geq 2$ is the simplest class to study by a very large margin. 

\m 

\n We next make use of Appendix B's criteria of graph theoretic nontriviality, noting for which $N$'s these first occur in Leibniz space.  
Two simplifying preliminaries to note are, firstly, removal of l cone points -- an $N$-independent feature for all but the very smallest $N$ -- 
creating a residue graph (Sec \ref{Residues}). 
This is relevant since the very smallest $N$'s scaled Leibniz spaces are just complete graphs $\mK_N$, so we are measuring a departure from this original trend.   
Secondly that almost immediately the complement of the residue is simpler to handle. 
This encodes which coincidence-or-collision notions obstruct all such being topologically adjacent to all others. 

\m 

\n A first set of simplicity criteria then pick out (Secs \ref{6,2}, \ref{6,S}, \ref{5,1}) $(\geq 2, 6)$, $(\mathbb{S}^1, 6)$ and (1, 5): 
these have a residue complement graph that is not just a collection of points, paths or cycles.  
These are the graphs displayed on the article's cover page figure \ref{Top-2-Front}. 
While all three of these are planar, the $(\mathbb{S}^1, 6)$ case's complement is nonplanar.   

\m 

\n Another simplicity criterion possessed by all of these residue complements and $(\geq 2, 7)$ and $(\mathbb{S}^1, 7)$ (Sec \ref{Sevens}) is modularity.  
This is taken to mean that the graph is a ladder of $\mK_p$ subgraph bock `rungs' each only attached to adjacent `rungs' along the ladder. 
This simplicity criterion, as well concurrent nonplanarity of the residue and its complement, 
first occur (Sec \ref{6,1}) for (1, 6), ($\geq$ 2, 8)
%
%
and ($\mathbb{S}^1$, 8). 
These last three graphs correspond to the most complicated topological Leibniz (scaled) shape spaces considered in the current article.  
Large-$N$ estimates for the number of vertices and edges are outlined in Sec \ref{Big-N}.  

\m 

\n Our Conclusion includes firstly comparison with metric-level scaled shape theory's orbit space genericity criterion \cite{ML00} which enters at $N = 5$. 
One idea is that consideration of rubber shapes gives a {\sl distinct} source of 6 and 8 body problem difficulties hitherto unremarked upon; 
see \cite{Minimal-N} for further discussion.
We secondly give a brief outline of a Topological Background Independence model \cite{I89, Witten89, GH92} 
and its metric-shape \cite{Top-Shapes} and General Relativity \cite{Mis-Top, Fischer70} counterparts.  

\m

\n Appendices on each of partitions, graphs and lattices are included; the last of these covers both partition refinement lattices, 
and the explanation of the particular significance of Leibniz spaces.   

\vspace{10in}

\section{Kendall-type Geometric Relational Theories}

\n{\bf Definition 1} {\it Carrier space} $\FrC^d$, alias absolute space in the physically realized case, is an at-least-provisional model for the structure of space.

\m

\n{\bf Remark 1} We restrict ourselves to carrier spaces which are connected manifolds without boundary.
Among these, the current article refers to $\FrC^d = \mathbb{R}^d$ -- the most common choice in the $N$-Body Problem and Shape Theory literature -- 
                                           $\mathbb{S}^1$, 
				 as well as to the general $\FrC^d$ of dimension $\geq 2$.
				 
\m

\n{\bf Remark 2} In some physical applications, the points model material particles (classical, and taken to be of negligible extent).
Because of this, we subsequently refer to constellations as consisting of points-or-particles. 

\m 

\n{\bf Definition 2} {\it Constellation space} is the product space 
\be 
\FrQ(\FrC^d, N) = \bigtimes_{i = 1}^N \FrC^d                \m ,    
\ee 
where $N$ is the number of points-or-particles under consideration.

\m 

\n{\bf Example 1} For $\FrC^d = \mathbb{R}^d$, the constellation space is 
\be 
\FrQ(\mathbb{R}^d, N) := \FrQ(d, N) = \bigtimes_{i = 1}^N \mathbb{R}^d = \mathbb{R}^{N \, d}                \m .      
\ee
\n{\bf Example 2} The other case considered in the current article is $\FrC^d = \mathbb{S}^1$: the circle, for which the constellation space is 
\be 
\FrQ(\mathbb{S}^1, N) = \bigtimes_{i = 1}^N \mathbb{S}^1 = \mathbb{T}^{N}                \m : \m \mbox{ the $N$-torus} \m .       
\ee
\n{\bf Structure 1} Relational Theory furthermore takes some group of automorphisms 
\be
\lFrg = Aut(\langle \FrC^d, \sigma \rangle)
\ee 
of $\FrC^d$ -- or $\FrQ(\FrC^d, N)$ by its product group structure -- and regards these as irrelevant to the modelling in question. 
$\sigma$ is here some level of mathematical structure on $\FrC^d$.  

\m

\n{\bf Example 1} Quotienting out the Euclidean group $Eucl(d)$ of translations and rotations in a bid to free one's modelling from $\mathbb{R}^d$ absolute space, 
as equipped with Euclidean metric structure.  
This is an example of quotienting out an isometry group, since 
\be
Isom(\mathbb{R}^d) = Eucl(d) \m .  
\ee 
\n{\bf Example 0} Quotienting out the similarity group $Sim(d)$ of translations, rotations and dilations so as to additionally be free of absolute scale. 
This is Kendall's choice for Shape Statistics \cite{Kendall84, Kendall89, Kendall}. 
It moreover turns out to be a useful intermediary and/or structure within Example 1's Mechanics context as well, 
whether in modelling whole universes \cite{RPM, AF, FileR, ABook} or subsystems as occur in Celestial Mechanics \cite{Smale70, M02-M05} or Molecular Physics. 
Because of this, our next mention of these two examples shall be in reverse order. 

\m

\n{\bf Example 2} For $\FrC^d = \mathbb{S}^1$, an isometry group --
\be 
Isom(\mathbb{S}^1) = SO(2) = U(1) = \mathbb{S}^1
\ee 
as a manifold -- exists. 
However, a distinct similarity group does not, since the generator of dilations does not respect the `periodic identification' of the circle.

\m 

\n{\bf Definition 3} {\it Relational space} is then the quotient space
\be  
\Rel(\FrC^d, N, Aut(\FrQ, \sigma)) \m = \m \frac{\FrQ(\FrC^d, N)}{Aut(\langle\FrQ, \sigma\rangle)}       \m .  
\ee
\n{\bf Definition 4} For those $\lFrg$ that do not include a scaling transformation, 
the relational space notion specializes to the {\it shape space} notion \cite{Kendall84, Kendall, FileR, AMech, PE16, A-Monopoles} 
\be 
\FrS(\FrC^d, N; \lFrg) := \Rel(\FrC^d, N; \lFrg)   \m . 
\ee
\n{\bf Definition 5} For those $\lFrg$ that do include a scaling transformation, 
the relational space notion specializes to the {\it shape-and-scale space} notion \cite{LR95, LR97, FileR, AMech, ABook, A-Monopoles} 
\be 
\FrR(d, N; \lFrg) := \Rel(\FrC^d, N; \lFrg)  \m .
\ee
\n{\bf Remark 2} Relational Theory is thus a portmanteau of Shape Theory and Shape-and-Scale Theory.  
The distinction of whether or not scaling is among the automorphisms is significant in practise because many of the most-studied models are part of a 
{\it shape space and shape-and-scale-space pair}.
This corresponds to Shape Theories which remain algebraically consistent upon removal of an overall dilation generator. 
However, there are more generally plenty of instances of singletons, of which one is given below and others are listed in e.g. \cite{ACirc, Project-1}.  

\m 

\n{\bf Example 0} For carrier space $\mathbb{R}^d$, quotienting out the constellation space by the similarity group $Sim(d)$ gives 
{\it Kendall's Similarity Shape Theory} \cite{Kendall84}.
In particular, in 1- and 2-$d$ for $\Gamma = id$, the shape spaces 
\be 
\FrS(d, \, N) := \FrS(\mathbb{R}^d, N; Sim(d))
\ee 
for this are the spheres $\mathbb{S}^{n - 1}$ in 1-$d$ and complex projective spaces $\mathbb{CP}^{n - 1}$ in 2-$d$, where $n :=  N - 1$. 

\m 

\n{\bf Example 1} For carrier space $\mathbb{R}^d$, quotienting out the constellation space by the Euclidean group $Eucl(d)$ gives {\it Metric Relational Theory}.
Some particular shape-and-scale spaces for this are
the real spaces $\mathbb{R}^{n}$ in 1-$d$ and cones over complex projective spaces $\mC(\mathbb{CP}^{n - 1})$ in 2-$d$.  

\m 

\n{\bf Example 2} The only further example we give here is for carrier space $\mathbb{S}^1$, because its Rubber Relational Theory is exceptionally its own universality class.  
Here quotienting out the constellation space by the isometry group $U(1)$ gives another metric shape-and-scale theory; 
the corresponding shape-and-scale space is 
\be 
\FrR(\mathbb{S}^1, N; U(1)) = \frac{\bigtimes_{i = 1}^N \mathbb{S}^1}{\mathbb{S}^1}  = \bigtimes_{i = 1}^n \mathbb{S}^1 \mathbb{T}^{n} \m : \m \mbox{ the $n$-torus} \m .  
\ee

\section{Discrete quotients and Leibniz spaces}

\n{\bf Remark 1} 
Quotienting by $\Gamma = id$, $C_2$-ref (acting reflectively), $S_N$ and $S_N \times C_2$ gives a first quartet of discrete quotients. 
These correspond to (no, no), (yes, no), (no, yes) and (yes, yes) answers to the twofold question of whether one's model possesses (mirror image, label) distinguishability.  
The `top discrete group' acting moreover collapses from $S_N \times C_2$ to just $S_N$ if $N$ is large enough relative to $d$ that mirror image identification 
becomes obligatory by rotation through extra dimensions to those spanned by the point-or-particle separation vectors.  
Including this discrete quotienting feature, we arrive at the following definition.  

\m 

\n{\bf Definition 1}  A {\it geometrical-level relational theory} is a quadruple 
\be
(\FrC^d, N; \lFrg, \Gamma)
\ee 
for $\FrC^d$ a carrier space, 
$N$ a point-or-particle number, 
$\lFrg$ a continuous group of automorphisms acting on $\FrC^d$, and 
$\Gamma$ a discrete group of automorphisms acting on $\FrQ(\FrC^d, N)$.
$\lFrg$ is in more detail $Aut(\langle\FrQ, \sigma\rangle)$, 
for $\sigma$ some level of mathematical structure on $\FrQ$ which is itself preserved by the automorphisms in hand.\footnote{This covers both $\lFrg$ and $\Gamma$. 
More generally, one could have an unsplittable group playing a joint role running over both of these.}

\m

\n{\bf Remark 2} Splitting into continuous and discrete automorphisms (where possible), 
\be
\Rel(\FrC^d, N; \lFrg, \Gamma) = \frac{  \Rel(\FrC^d, \, N)  }{  \lFrg \circ \Gamma  }   \m .
\ee
Here $\circ$ is a generic product (of the form $\times$ or $\rtimes$ -- semidirect product of groups -- in all examples in the current article).

\m 

\n{\bf Remark 3} Within this scheme,  
\be 
\Rel(\FrC^d, N; id , id)   \es  \frac{\FrQ(\FrC^d, N)  }{  id \times id  } 
                           \es  \frac{\FrQ(\FrC^d, N)  }{  id  }
                           \=:  \FrQ(\FrC^d, N)                             \m :  
\ee
constellation space itself.  

\m

\n{\bf Remark 4} If one quotients by the top discrete group Leibniz space ensues: 
a fitting name in connection with Leibniz' Identity of Indiscernibles out of this space containing precisely 1 of every type of configuration.  
Considering this removes symmetry from contention in Relational Theory: passing from a tessellation to a single tile: the Leibniz space.  

\m

\n{\bf Remark 5} For $N \geq 3$, moreover, partial label distinguishabilities exist, extending the above list of four possibilities.
Another conceptually and technically useful way \cite{A-Monopoles} of viewing this extension concerns the lattice of distinguishable group actions of the subgroups of $S_N \times C_2$ 
or $S_N$. 
See \cite{A-Monopoles, Top-Shapes} for the intervening subgroup and subgroup action lattices between these and $id$ for small $N$.

\section{Rubber shapes and their Topological Shape Theory}\label{Rubber}

\n{\bf Definition 1} The {\it rubber shape} alias {\it topological notion of shape} is the topological content of normalizable $N$-point constellations 
on some carrier space $\FrC^d$.  
If one drops the normalizability condition, the maximal coincidence-or-collision O can also be included; in this case one is dealing rather with scaled shapes.  

\m 

\n{\bf Notation 1} We depict topological (scaled) shapes by `topological distribution diagrams' in pastel sky blue.  
This colouring renders them immediately distinguishable from 
                                                     geometrical (scaled) shapes              (grey) 
                                                  and geometrical relational spaces    (black).   

\m 

\n{\bf Remark 1} This is a fairly weak notion of (scaled) shape since many of the properties usually attributed to $N$ point constellations are metric in nature: 
the angles defining an isosceles triangle, the ratios defining a rhombus...
As Fig \ref{Top-Leib-1-to-4-2}'s example will show, the topological notion of shape is not however empty.
Topological shapes furthermore provide useful insights as regards the structure of relational spaces of geometric-level configurations.

\m 

\n{\bf Definition 2} The above admit a considerably simpler relational theory: {\it Rubber Relational Theory} alias {\it Topological Relational Theory}, which is characterized by just 
\be
(D, N; S, \Gamma) \m . 
\ee 
$S$ is here binary-valued: without and with scale: $\emptyset$, and $s$ standing for `scaled'. 

\m 

\n $D$ is ternary-valued: 1 or $\geq 1$, with the first case split furthermore into distinct open $\mathbb{R}^1$ and closed $\mathbb{S}^1$ cases.  
We denote the circle case by $1^{\prime}$ and the other two cases collectively by $d$.  

\m 

\n{\bf Remark 2} N.B.\ also what this does not depend on: neither a choice of $d \geq 2$ carrier space, 
nor a choice of continuous automorphism group aside from whether scale is included. 
The reason that whether a model possesses scale survives as a difference  
is through the aforementioned distinction of whether to include the maximal coincidence-or-collision O.  

\m  

\n{\bf Definition 3} The {\it topological relational space}\footnote{Topological relational spaces are to be distinguished in this article in writing by the preface $\FrT$op.} 
\be
\Top\mbox{-}\Rel(D, N; S, \Gamma)                                                           \m .
\ee
is the configuration space concept encompassing both rubber shapes, for which one has a {\it topological shape space},  
\be
\Top\mbox{-}\FrS(D, N; \Gamma) =  \Top\mbox{-}\Rel(D, N; \emptyset, \Gamma)             \m ,
\ee
and the rubber scaled shapes, for which one has a {\it topological scaled-shape space},  
\be
\Top\mbox{-}\FrR(D, N; S, \Gamma)  = \Top\mbox{-}\Rel(D, N; s, \Gamma)                  \m .  
\ee
\n{\bf Definition 4} A given model's topological (scaled) shapes are moreover interlinked by the following {\it topological adjacency condition}.  
At a purely topological level, two topological (scaled) shapes are topologically adjacent if the points-or-particles of one can be obtained by a single step's worth of 
fusion or fission. 
The single step is defined as carrying out as many fissions as one pleases, or as many fusions as one pleases, but not a mixture of both.  

\m

\n{\bf Remark 3} In particular, this means that one cannot fission a point-or-particle out of one coincidence-or-collision and then also fuse it into a distinct 
coincidence-or-collision in the same step.

\m 

\n{\bf Structure 1} Topological relational spaces are graphs, with topologically distinct rubber (scaled) shapes as vertices and topological adjacency relations encoded by the edges. 

\m 

\n{\bf Notation 2}  We depict these topological relational space graphs in bright blue, so as to distinguish them from the above pastel blue, grey and black levels of structure, 
and also from other or abstract uses of graphs (bright purple).  

\m 

\n{\bf Remark 4} In the geometrical shape version, the above definition of topological adjacency 
coincides with whether one topologically distinct class of (scaled) shapes is realized as an edge, corner... of another such (Fig \ref{Top-Adj}).  
Because of this, topological relational space graphs already encode some features of how geometric relational spaces for the same $(d, N, \Gamma)$ fit together. 
This is with the added bonus that with increasing $d$ and/or $N$, 
geometrical relational spaces rapidly become high-dimensional manifolds with large networks of lower dimensional pieces: corners, edges and generalizations. 
Or indeed high-dimensional stratified manifolds with numerous lower-dimensional strata.  
So it is useful that at least the topological (scaled) shape graph information remains readily calculable 
[and readily depictable, at least for the first few $N$'s and $d$'s for which the geometrical (scaled) shape space has ceased to be readily depictable].  
%
%
{            \begin{figure}[!ht]
\centering
\includegraphics[width=0.8\textwidth]{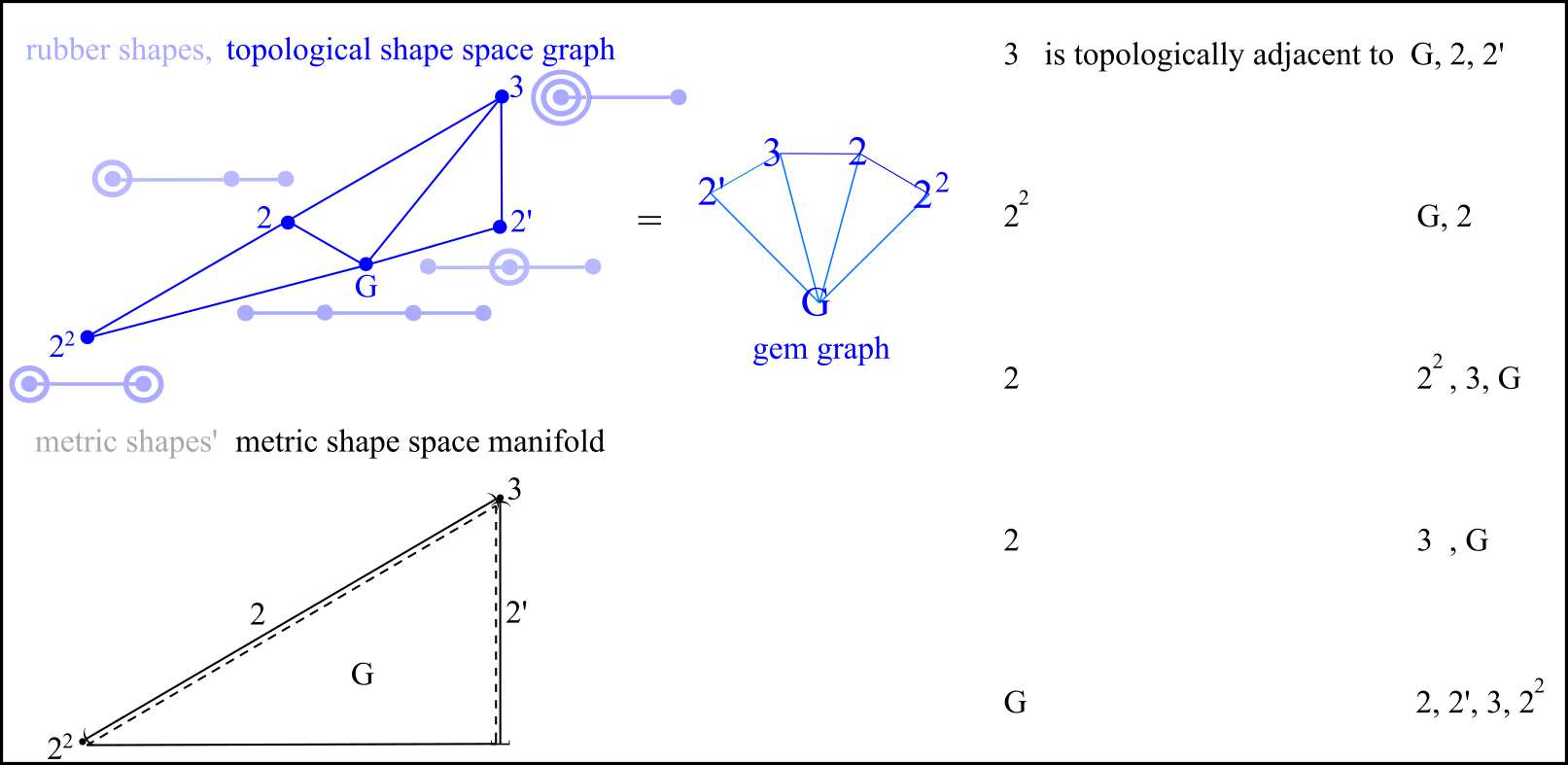}
\caption[Text der im Bilderverzeichnis auftaucht]{        \footnotesize{Topological adjacency at the level of rubber (scaled) shapes.
and of geometrical (scaled) shapes' configuration space, for the (1, 4) model example.} }
\label{Top-Adj} \end{figure}          }

\m

\n{\bf Remark 5} There is moreover a general result for how the scaled case's maximal coincidence-or-collision is appended to the corresponding topological shape space graph. 

\m 

\n{\bf Lemma 1}  If $\mG$ is a given topological shape theory's shape space, 
then the cone graph $\mC(\mG)$ is the corresponding topological scaled shape theory's scaled shape space. 
The cone vertex here is the maximal coincidence-or-collision O.

\section{Previous study of ($\geq$ 2, $\biN$) class: partitions}\label{Partitions}

\n{\bf Proposition 1} In this class, rubber scaled shapes are in 1 : 1 correspondence with partitions. 

\m 

\n{\bf Remark 1} This is a useful observation, since partitions are more established mathematical problem as per Appendix A.  

\m 

\n{\bf Remark 2} See Fig \ref{Top-Leib-1-to-4-2} for the corresponding $N$ = 1 to 4 topological relational space graphs.  
Scaled and unscaled cases are different, in terms of appending the maximal collision.  
This is the first place that the maximal coincidence-or-collision makes a difference, 
but it does not present a difficulty on this occasion: it is just a vertex like any other partition. 
%
{            \begin{figure}[!ht]
\centering
\includegraphics[width=1.0\textwidth]{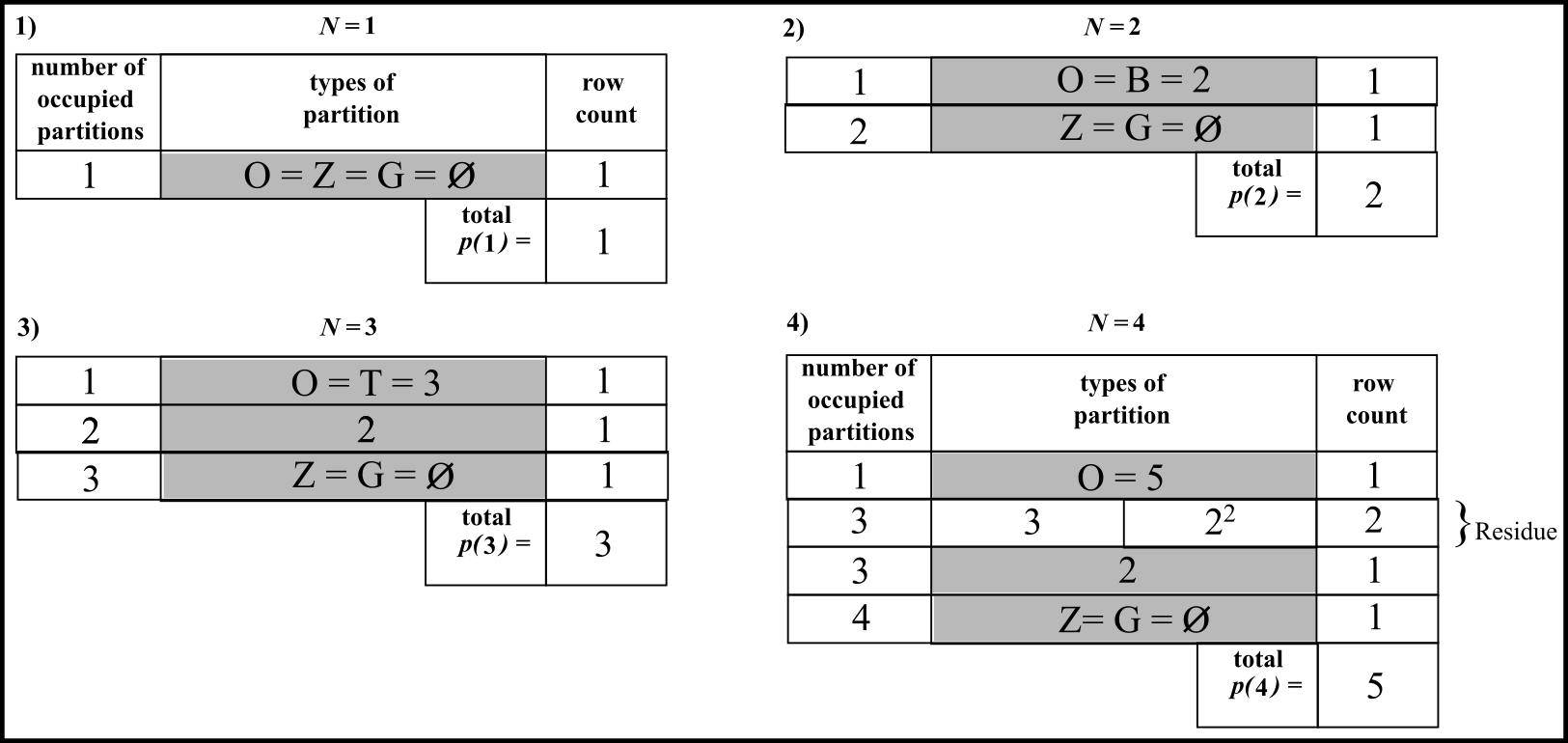}
\caption[Text der im Bilderverzeichnis auftaucht]{        \footnotesize{Topological configurations for $N = 1$ to 4.
} }
\label{Small-Table} \end{figure}          }
%
{            \begin{figure}[!ht]
\centering
\includegraphics[width=0.65\textwidth]{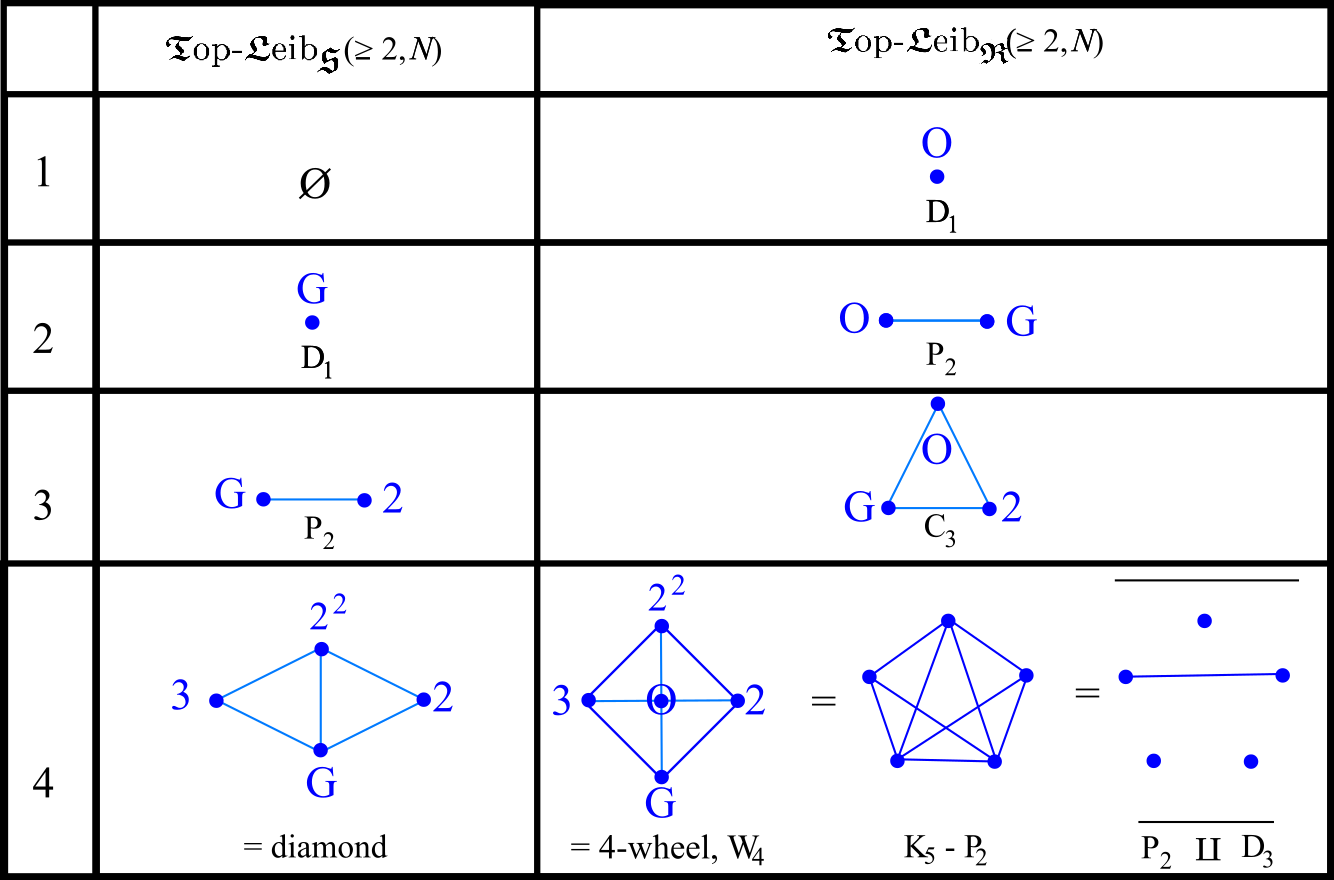}
\caption[Text der im Bilderverzeichnis auftaucht]{        \footnotesize{Topological Leibniz space graphs for $N = 1$ to $4$ in $d \geq 2$ \cite{I, III, Top-Shapes, IV}. 
} }
\label{Top-Leib-1-to-4-2} \end{figure}          }

\m 

\n{\bf Notation} In previous papers, I used G, B, T, Q and O for, respectively, generic configurations, and binary, ternary, quaternary and maximal coincidence-or-collisions.   
Partition-theoretic notation however becomes very quickly becomes more efficient with increasing $N$. 
In the single-occupancy-partitions removed version of this (Appendix A), the above are -, 2, 3, 4 and N (for $N \geq 2$). 
G and O moreover retain {\it lattice-theoretic} privilege as well as being the subject of most discussions in the $N$-Body Problem and Shape Theory, 
so these retain the right to be accorded special symbols. 
We furthermore take O to stand for the lattice-theoretic `one' (Appendix C) here, by which it sometimes makes sense to refer to G as the lattice-theoretic counterpart Z for zero.  
This helps in various other ways: statements about any $N$'s maximal collision on the one hand, and - being a rather confuseable symbol especially in copious use.  
So our default symbols are the lattice top-and-bottom privileged, but elsewise single-occupancy-partitions removed, G = Z, 2, 3, ..., N = O, 
alongside powers and strings of non-extremal lattice members of the `middle'. 
So the double-binary $\mB^2$ is $2^2$ in this notation, and the ternary-and-binary $\mT\mB$ is $3 \, 2$.

\section{Partition Theory is not enough for Shape Theory}\label{Not-Enough}

\subsection{Partition Refinement is not enough for topological adjacency}

\n{\bf Remark 1} Arranging the rubber (scaled) shapes, via their 1 : 1 correspondence with the partitions, 
as the partition refinement lattice $\lattice_{\sP}(N)$ of Appendix C, turns out to be useful.  
This is moreover not quite what we require for topological adjacency in Leibniz space, $\lattice_{\sP}(N)$ encoding rather dimensional descent one dimension at a time.

\m 

\n{\bf Example 1} These two concepts are already distinct for $N = 3$ since corners can be adjacent to faces as well as to edges.

\m 

\n{\bf Difference 1} thus involves topological adjacency in general encoding graph edges between graph vertices that more than one floor apart. 

\m 

\n{\bf Difference 2} is that topological adjacency is an undirected quantity, so the refinement lattice's arrows are rendered moot.

\subsection{Partitions are not enough for $\FrC^d = \mathbb{R}$ or $\mathbb{S}^1$ configurations}

\n{\bf Remark 1} The $N = 1, 2, 3$ cases of these coincide with their $d \geq 2$ counterparts \cite{I, II, Top-Shapes}. 

\m 

\n{\bf Proposition 1} The first distinction for $\mathbb{R}$ occurs for $N = 4$, differing due to the $2$ and $2^{\prime}$ distinction (Fig \ref{(4,1)-Summary}.a).
This does not affect $(\mathbb{S}^1, 4)$ by cycling these topological configurations into each other (Fig \ref{Top-Leib-6-S-Table}.a).  
%
{            \begin{figure}[!ht]
\centering
\includegraphics[width=0.85\textwidth]{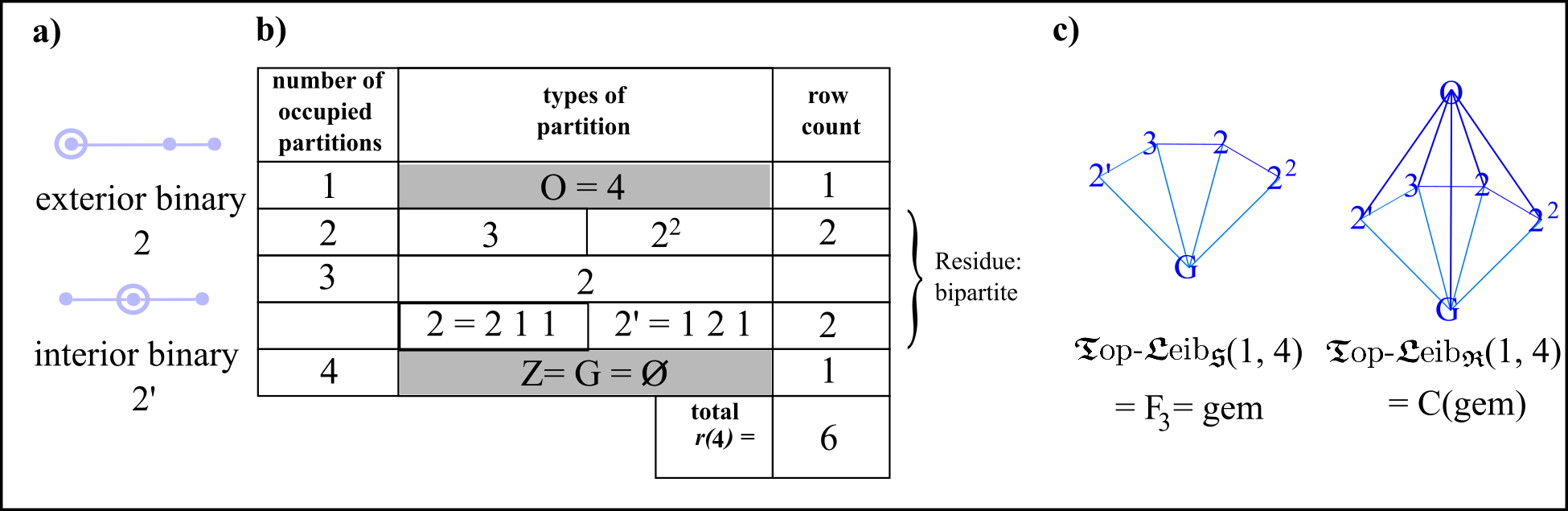}
\caption[Text der im Bilderverzeichnis auftaucht]{        \footnotesize{$(4, \mathbb{R})$'s topological configurations.  
Cone and residue features marked on such figures shall be explained in Secs 7 and 8.} }
\label{(4,1)-Summary} \end{figure}          }

\m 

\n{\bf Remark 2} This leads to a further combinatorial problem, of evaluating 
\be 
s(N) := \#(    \mbox{ rubber scaled shapes configurations on }  \mathbb{S}^1    )  \m . 
\ee 
\n{\bf Structure 1} The above distinction can moreover be interpreted as a refinement in excess of those of the partition refinement lattice, 
splitting Fig \ref{Top-Leib-1-to-4-2}.4)'s configurations into Fig \ref{(4,1)-Summary}.b)'s.   

\m 

\n{\bf Proposition 2} The corresponding topological Leibniz spaces were given in \cite{II, Top-Shapes}; it is presented in Fig \ref{(4,1)-Summary}.c) for sake of self-containedness.  

\m 

\n{\bf Remark 3} $N = 4$ does not suffice for $\mathbb{S}^1$ to exhibit a refinement. 
Nor does $N = 5$, for which the joint situation in $\geq 2$ and $\mathbb{S}^1$ is given in Fig \ref{Top-Leib-5-2-Table}.  
%
{            \begin{figure}[!ht]
\centering
\includegraphics[width=0.45\textwidth]{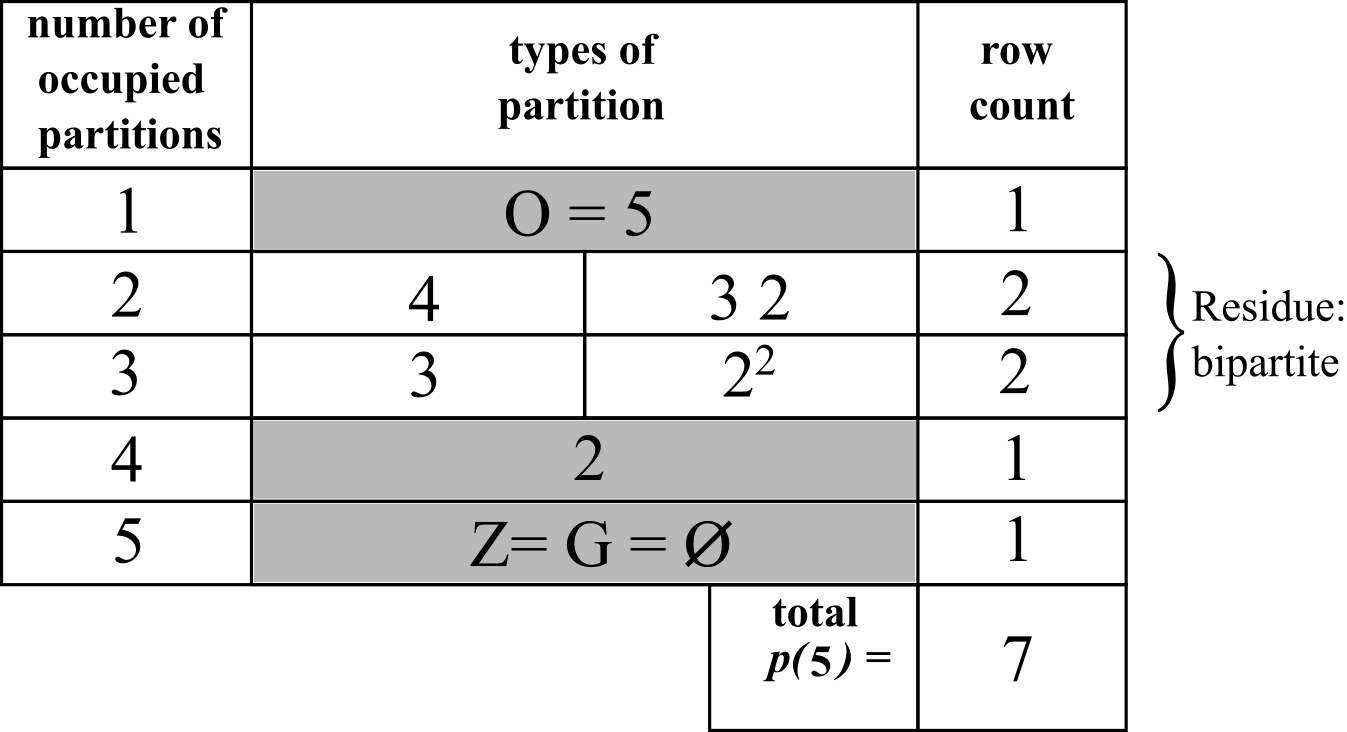}
\caption[Text der im Bilderverzeichnis auftaucht]{        \footnotesize{$(\geq 2, 5)$'s 7 partitions of 5 indistinguishable objects.} }
\label{Top-Leib-5-2-Table} \end{figure}          }

\m 

\n{\bf Remark 4} $N \geq 6$ does however suffice for such a distinction, 
based on Fig \ref{Top-Leib-6-S-Table}.b)'s rubber configurations splitting $\geq 2$'s analysis of Fig \ref{Top-Leib-6-2-Table} into Fig \ref{Top-Leib-6-S-Table}.c)'s $\mathbb{S}^1$ analysis.
This leads to a yet further combinatorial problem, now of evaluating  
\be 
s(N) := \#(    \mbox{  rubber scaled shape configurations on } \mathbb{S}^1    )  \m . 
\ee 
%
{            \begin{figure}[!ht]
\centering
\includegraphics[width=0.45\textwidth]{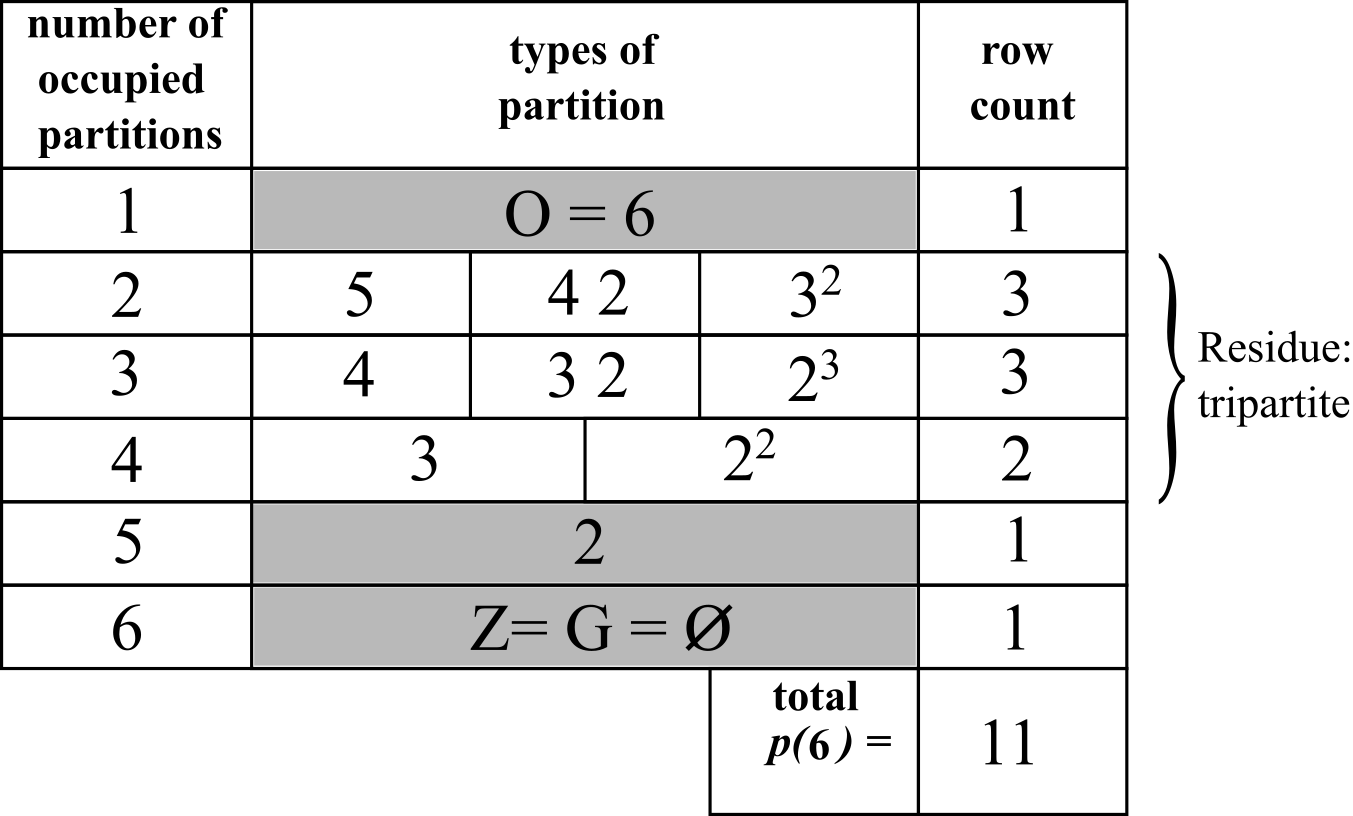}
\caption[Text der im Bilderverzeichnis auftaucht]{        \footnotesize{$(\geq 2, 6)$'s 11 partitions of 6 indistinguishable objects.} }
\label{Top-Leib-6-2-Table} \end{figure}          }
%
{            \begin{figure}[!ht]
\centering
\includegraphics[width=0.9\textwidth]{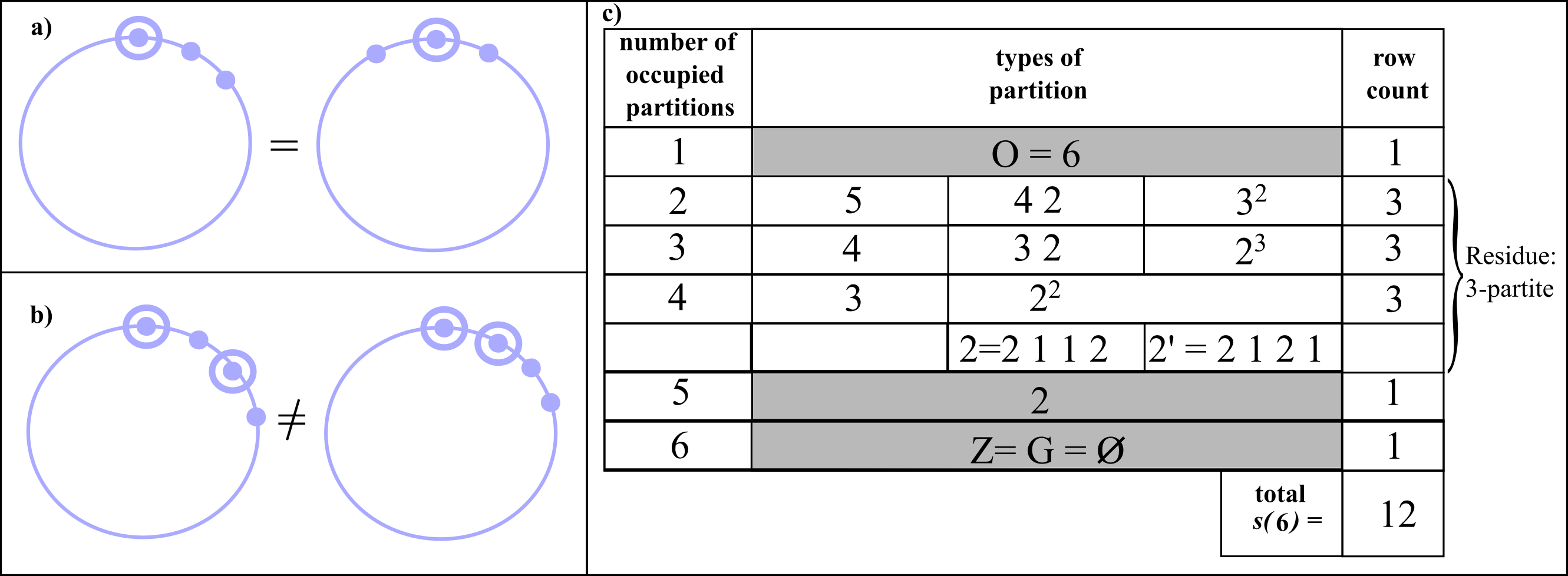}
\caption[Text der im Bilderverzeichnis auftaucht]{        \footnotesize{a) and b) show how 6 rather than 4 points-or-particles are required to incurr a departure from 
the 1 : 1 correspondence with partition functions on the carrier space $\mathbb{S}^1$.    
c) $(\mathbb{S}^1, 6)$'s 12 distinct rubber configurations.
} }
\label{Top-Leib-6-S-Table} \end{figure}          }

\m 

\n{\bf Proposition 3} These configurational distinctions are successive refinements, by which we have the bounds 
\be 
r(N)  \geq  s(N) 
      \geq  p(N)   \m .
\label{RSP}
\ee  
\n The $\mathbb{R}$ class is by far the hardest to work with; see Fig \ref{Top-Leib-5-1-Table} for the $N = 5$ case's rubber configurations.    
%
{            \begin{figure}[!ht]
\centering
\includegraphics[width=0.55\textwidth]{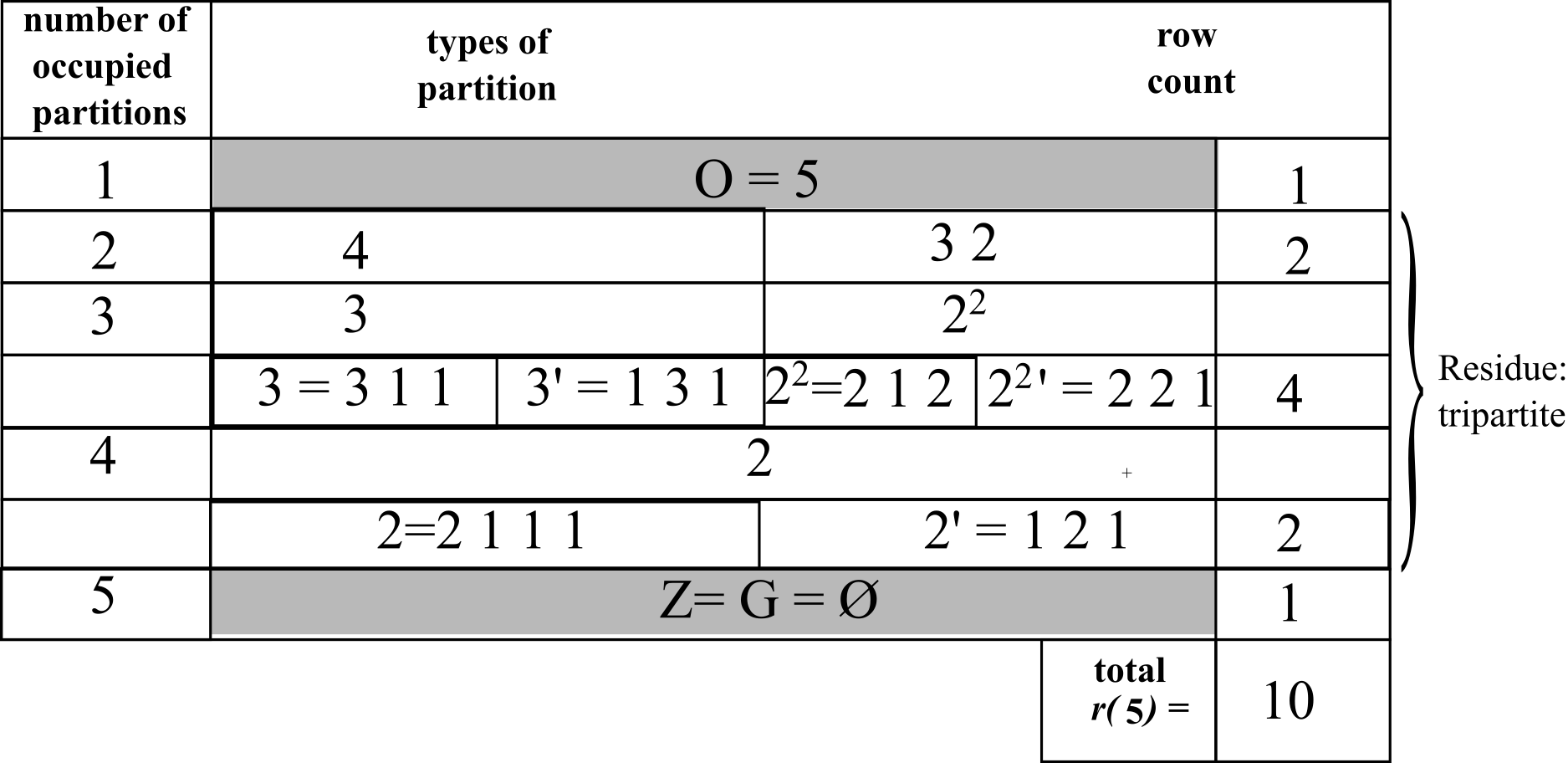}
\caption[Text der im Bilderverzeichnis auftaucht]{        \footnotesize{(1, 5)'s 10 distinct rubber configurations.
} }
\label{Top-Leib-5-1-Table} \end{figure}          }
 
\section{Cone graphs in Topological Relational Theory}\label{Cones}

\n{\bf Remark 1} Observing the following pattern in Figs \ref{Top-Leib-1-to-4-2} and \ref{(4,1)-Summary} turns out to be useful for any further $N$. 
\be 
\mK_1 = \mC(\emptyset)                               \m , 
\ee 
\be 
\mK_2 = \mC(\mK_1) = \mC^2(\emptyset)                \m ,
\ee 
\be 
\mK_3 = \mC(\mK_2) = \mC^2(\mK_1) = \mC(\emptyset)   \m ,
\ee 
\be 
\mbox{diamond} = \mC(\mP_3) =\mC^2(\mD_2)            \m , 
\ee 
\be 
\mC(\mbox{diamond}) = \mC^2(\mP_3) =\mC^3(\mD_2)     \m ,    
\ee
where $\mC$ denotes cone over a graph as per Appendix B.2.  
Thus all the topological Leibniz spaces considered so far are cones. 

\m 

\n{\bf Remark 2} It is moreover straightforward to establish that 
\be 
\mbox{ for \m $N \geq 2  \mma$ all $\Top\mbox{-}\Leib_{\sFrS}(\geq 2, N)$ \m are double cones with cone points G, B, and }   
\ee 
\be 
\mbox{ for \m $N \geq 3 \mma$ all $\Top\mbox{-}\Leib_{\sFrR}(\geq 2, N)$ \m are triple cones with cone points G, B, and O}  \m .   
\ee
\n{\bf Remark 3} Cone graphs also feature as shape spaces for $d \geq 2$ because in these cases the generic configuration $\mG$ is topologically adjacent to 
all other configurations.  

\m 

\n{\bf Remark 4} Cones over cones of graphs, $\mC(\mC(\mG))$, also feature in the corresponding $d \geq 2$ scaled shape spaces, with O and G as cone points (in either order).

\m 

\n In fact, two further objects of interest from the study of Shape Theory/$N$-Body Problem are further such cones. 

\m 

\n{\bf Definition 1} The {\it coincidence-or-collision structure} of scaled topological Leibniz space, $\Top\mbox{-}\Co\mbox{-}\Leib_{\sFrR}(\FrC^d, N)$ 
is the set of rubber shapes involving at least one coincidence-or-collision;  
\be
\mbox{ for $N \geq 2 \mma$ all \mbox{ } $\Top\mbox{-}\Co\mbox{-}\Leib_{\sFrR}(\geq 2, N)$ \mbox{ } are double cones with cone points B and O}  \m . 
\ee 
\n{\bf Definition 2} The {\it coincindence-or-collision structure} of pure-shape topological Leibniz space, $\Top\mbox{-}\Co\mbox{-}\Leib_{\sFrS}(\FrC^d, N)$ 
is the set of {\sl normalizable} rubber shapes involving at least one coincidence or collision;  
\be
\mbox{ for \m $N \geq 2$ \mma all $\Top\mbox{-}\Co\mbox{-}\Leib_{\sFrS}(\geq 2, N) $ are cones with cone point B}  \m . 
\ee 
\n So what we really want to compute are `residues' that the other four objects above are the $\mC_{2}$, $\mC^2_{\sO, 2}$, $\mC^2_{2, \sG}$ and $\mC^3_{\sO, 2, \sG}$ over.  

\m 

\n{\bf Definition 3} The {\it residue} $\Res(\mG)$ of a cone graph $\mG = \mC^k(\mH)$ is the end product of identifying and removing all cone points. 
We use in particular $\Res(\FrC^{d}, N)$ as shorthand for $\Res(\Top\mbox{-}\Leib(\FrC^d, N)$).  

\m 

\n{\bf Remark 5} This insight cuts down our incipient table to just Fig \ref{Res-1-to-4-2}. 
It is also at this point that we pass to our definitive notation. 
The first point of the current paper is that this so far very trivial table very quickly becomes graph-theoretically nontrivial with increase in $N$.  
%
{            \begin{figure}[!ht]
\centering
\includegraphics[width=0.25\textwidth]{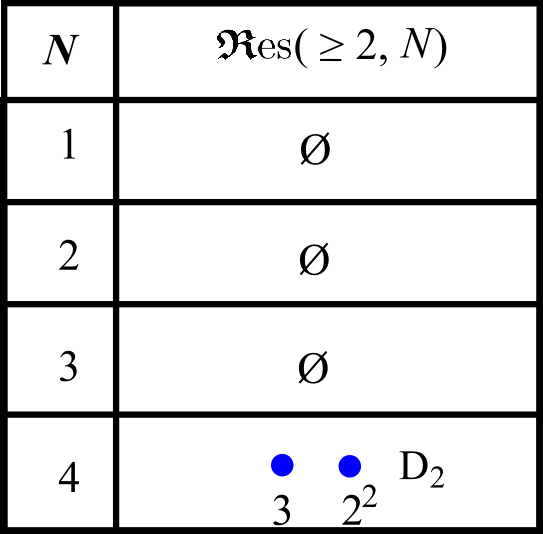}
\caption[Text der im Bilderverzeichnis auftaucht]{        \footnotesize{Residues for $N = 1$ to $4$ in dimension $\geq 2$.} }
\label{Res-1-to-4-2} \end{figure}          }

\m

\n{\bf Remark 6} The current section moreover justifies Appendix B.2 and 5's outlay of cone graph concepts and lemmas.

\m 

\n{\bf Proposition 1} For $(1, N \geq 4)$, 2 is no longer a cone point. 
This means that $\Res = \Co_{\sFrS}$, and everything else has one less cone point.   

\m 

\n{\bf Remark 7} (1, 4)'s $\Res$ is still however trivial, as per Fig \ref{(4,1)-Summary}.  
While $(\mathbb{S}^1, 5)$ also remains indistinguishable from $(\geq 2, 5)$, 
$N = 6$ suffices for $\mathbb{S}^1$ to $\geq 2$ distinction.
The only difference here is than just $\Co_{\sFrR}$ and $\Leib_{\sFrR}$ are relational-theoretically well defined cones over the residues. 

\m 

\n{\bf Remark 8} Let us end this section with Fig \ref{Top-Space-Graph-Order}'s table of counts.
%
{            \begin{figure}[!ht]
\centering
\includegraphics[width=0.85\textwidth]{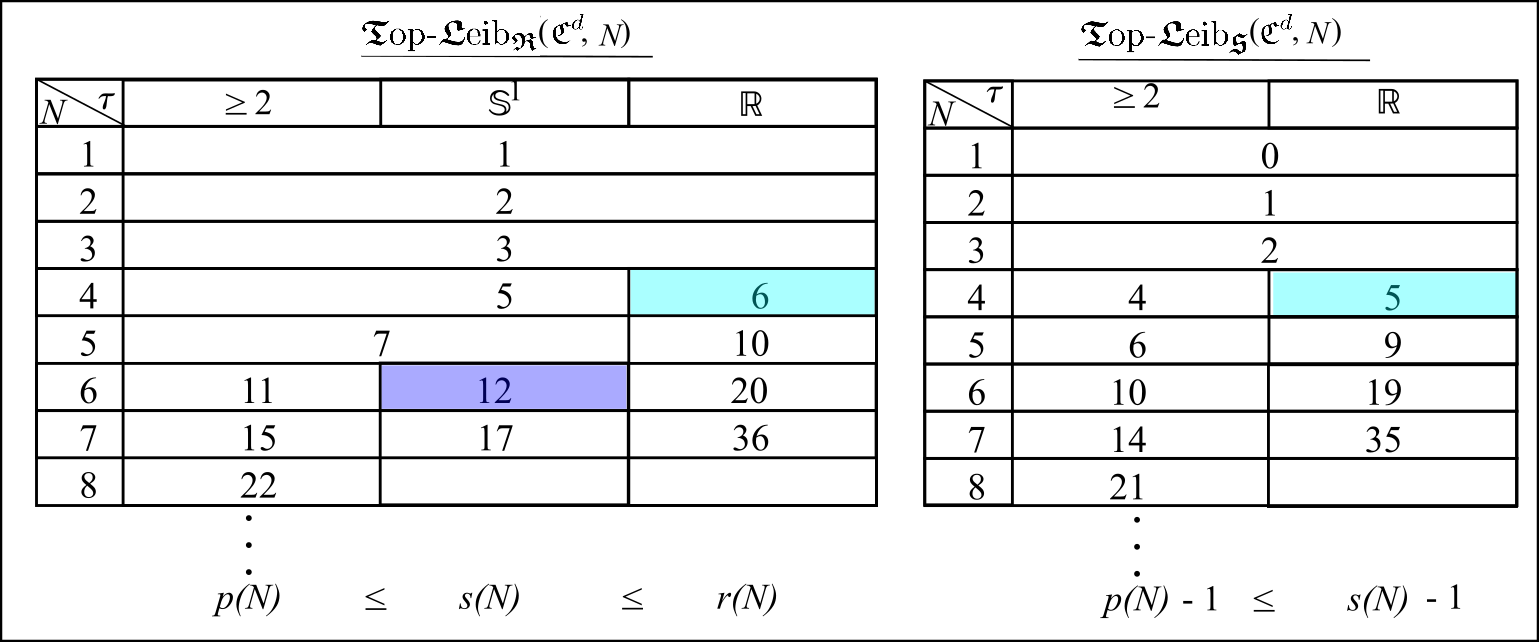}
\caption[Text der im Bilderverzeichnis auftaucht]{        \footnotesize{$(d, N)$ grid of orders of topological Leibniz space graphs. } }
\label{Top-Space-Graph-Order} \end{figure}          }

\section{Graphically-nontrivial Leibniz space residues}\label{Residues} 

\n We begin to address this with some simple nonrepresentability results.

\m 

\n{\bf Non-representability Lemma 1} Not all graphs are Leibniz spaces.  

\m

\n\underline{Proof} 
Not all graphs are double cones.  
The smallest example which is not is $\mD_2$, and the smallest connected example which is not is $\mP_3$.    
$\Box$ 

\m 

\n{\bf Remark 1} Thus Leibniz space topological graphs are a strict subset of graphs. 

\m 

\n{\bf Non-representability Lemma 2} There are numbers for which no graph of that order is a topological Leibniz space. 

\m

\n\underline{Proof} 
By successive fine grainings, (\ref{RSP}) holds.   
So if value $g_0$ has not yet occurred by the row ending in $p(N)$, it cannot occur further down the table.  

\m 

\n For shape spaces, this returns the minimal 
\be 
g_0^{\sFrS} = 3                \m ,
\ee 
and for scaled shape spaces, the minimal 
\be
g_0^{\sFrR} = 4                \m ;
\ee
for the two together, 
\be 
g_0 = 8                        \m .
\ee   
$\Box$ 

\m 

\n{\bf Remark 2} The above three counterexamples require checking up to $N = 4$, 4 and 6 respectively, thus constituting 4- and 6-body problem results.

\m  

\n{\bf Non-representability Lemma 3} Not all graphs are topological shape spaces (with reference to connected manifolds without boundary in the role of carrier spaces). 

\m 

\n\underline{Proof} 
Use 
\be 
|\FrS(\FrC^d, N; S, \Gamma)| \geq  |\Top\mbox{-}\Leib_{\sFrS}(\FrC^d, N)| 
\ee 
to truncate the necessary search. 

\m

\n Observe that both order-2 graphs occur, whereas for order 3, $\mP_3$ and $\mC_3$ do, but $\mD_3$ and $\mP_2 \, \disjoint \,  \mD_1$ do not. 

\m

\n If only [graphs] -- graphs modulo complementation -- are considered, this example does not suffice because the excluded graphs are the included graphs' complements. 

\m

\n Even by this strengthened criterion, however, $g_0 = 4$ suffices, 
since only $\mP_4$, $\mC_4$, diamond and claw (Fig \ref{Graph-6}) are realized as topological shape spaces \cite{Top-Shapes}, 
leaving the `$\mK_4$, $\mD_4$' and `paw, co-paw' complement pairs unrealized.                                                                                          
$\Box$ 

\m 

\n{\bf Remark 3} This is similarly a 4-body problem result, as the largest-$N$ item that requires checking occurs as $\Top\mbox{-}\Leib_{\sFrS}(4, \FrC^{\geq 2})$ 
(which just returns diamond again, this having already ocurred in a different labelling for $N = 3$).  

\m 

\n This is based on no graphs of a given order  having a property meaning that no topological Leibniz space of that order can have that property either.

\m 

\n The reader should now turn to Appendix B.3's graphical nontriviality criteria.

\section{First graphical nontriviality in the $\geq 2$ class}\label{6,2}

\n{\bf Remark 1} The $N = 1$ pure-shape case is trivial by criterion 0: it is the empty graph. 

\m 

\n The $N = 1$ scaled case and the $N = 2$ pure-shape case are trivial by criterion 1: these are $\mD_1$ graphs. 

\m 

\n The $N = 2$ scaled case and $N = 3$ pure-shape case are trivial by criterion 2: these are path graphs. 

\m 

\n The $N = 3$ scaled case is trivial by criterion 3: this is a cycle graph. 

\m 

\n The $N = 4$ pure-shape case is the diamond graph, whose complement is a collection of $\mD_1$'s and paths (Fig \ref{Graph-6}.a) and thus trivial by criteria 4 and 5.
Alternatively, diamond is the double cone over $\mD_2$, and so trivial by criterion 5. 

\m 

\n The $N = 4$ scaled case is the 4-spoked wheel graph $\mW_4$, which is the cone over the cycle $\mC_4$ and so trivial by criterion 6. 
Alternatively, its complement is $\mD_1$ from the cone point and two $\mP_2$'s: the complement of the square (Fig \ref{Graph-6}), which is trivial by criteria 4 and 5.  

\m 

\n Given criterion 6's triviality of cones of trivial graphs, we need only consider the pure-shape case since passing to the scaled case by addition of the 
maximal coincidence-or-collision O as a cone point preserves triviality status.  

\m 
 
\n{\bf Proposition 1} For $N = 5$ we split up the topological configurations as per Fig \ref{Top-Leib-5-2-Table}.a).  
Since the top row and bottom two rows are just conings, the remaining content is in the bipartite graph of Fig \ref{Res-5-2}.a).
This can furthermore be trivially straightened into 
\be 
\Top\mbox{-}\Leib_{\sFrS}(2, 5)  =  \mC_4 \m  \mbox{ labelled as per Fig \ref{Res-5-2}.b) } \m 
                                \es \overline{\mP_2 \, \disjoint \,  \mP_2} \m \mbox{ labelled as per Fig \ref{Res-5-2}.d)}  \m .
\ee 								
{            \begin{figure}[!ht]
\centering
\includegraphics[width=0.85\textwidth]{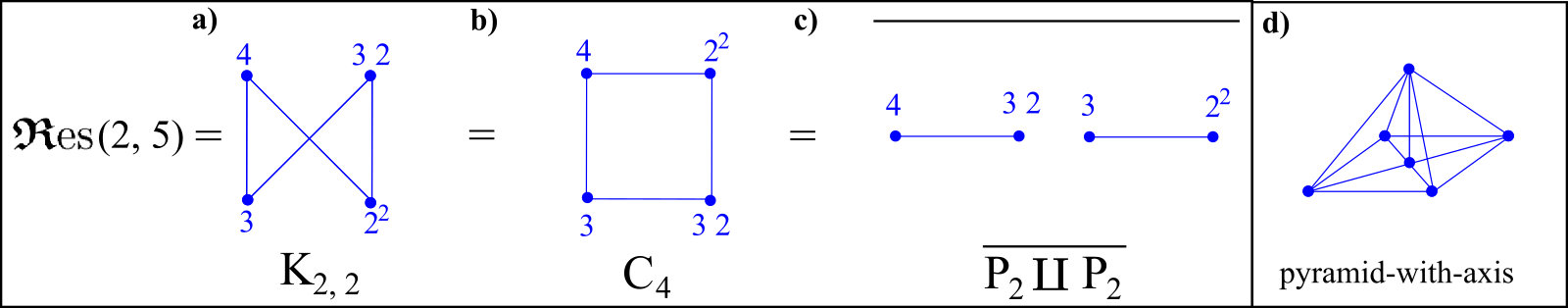}
\caption[Text der im Bilderverzeichnis auftaucht]{        \footnotesize{$\Res(\geq 2, 5)$ in a) bipartite, b) straightened out and c) complement forms.
d) is required for the double cone thereover.} }
\label{Res-5-2} \end{figure}          }

\m  

\n Thus 
\be 
\Res(\geq 2, 5)         =  \mC_4 
		   \es  \overline{\mP_2 \, \disjoint \,  \mP_2}                                      \m . 
\ee  
{\bf Corollary 1} From this, we can read off 
\be 
\Top\mbox{-}\Co\mbox{-}\Leib_{\FrS}(\geq 2, 5)  =  \mC_{\sB}(\Res(\geq 2, 5)) 
                                  =  \mC(\mC_4) 
						          =  \mW_4 
                                 \es  \overline{\mP_2 \, \disjoint \,  \mP_2 \, \disjoint \,  \mD_1}   \m , 
\ee  
\be 
\Top\mbox{-}\Leib_{\FrS}(\geq 2, 5)  =  \mC^2_{\sB, \sG}(\Res(\geq 2, 5))
                              =  \mC^2(\mC^4)
                              =  \mC(\mW_4)
                              =  \mbox{pyramid-with-axis} 							  
                             \es  \overline{\mP_2 \, \disjoint \,  \mP_2 \, \disjoint \,  \mD_2}       \m , 
\ee  
with $\Top\mbox{-}\Co\mbox{-}\Leib_{\FrS}(\geq 2, 5)$ isometric to the previous but with O instead of G as second cone point, and finally  
\be 
\Top\mbox{-}\Leib_{\FrR}(\geq 2, 5)  =  \mC^3_{\sO, \sB, \sG}(\Res(\geq 2, 5))
                                     =  \mC^3(\mC_4)
                                     =  \mC(\mbox{pyramid-with-axis})   							  
                                    \es  \overline{\mP_2 \, \disjoint \,  \mP_2 \, \disjoint \,  \mD_3}  \m .  
\ee
See Fig \ref{Graph-2} for the 4-wheel graph $\mW_4$ and Fig \ref{Res-5-2}.d) for the pyramid-with-axis graph.    
These are all of course still trivial graphs according to criteria 1 to 5. 
 
\m 

\n{\bf Remark 2} We next split up $(\geq 2, 6)$'s topological configurations as per Fig \ref{Top-Leib-6-2-Table}.  
%
{            \begin{figure}[!ht]
\centering
\includegraphics[width=0.8\textwidth]{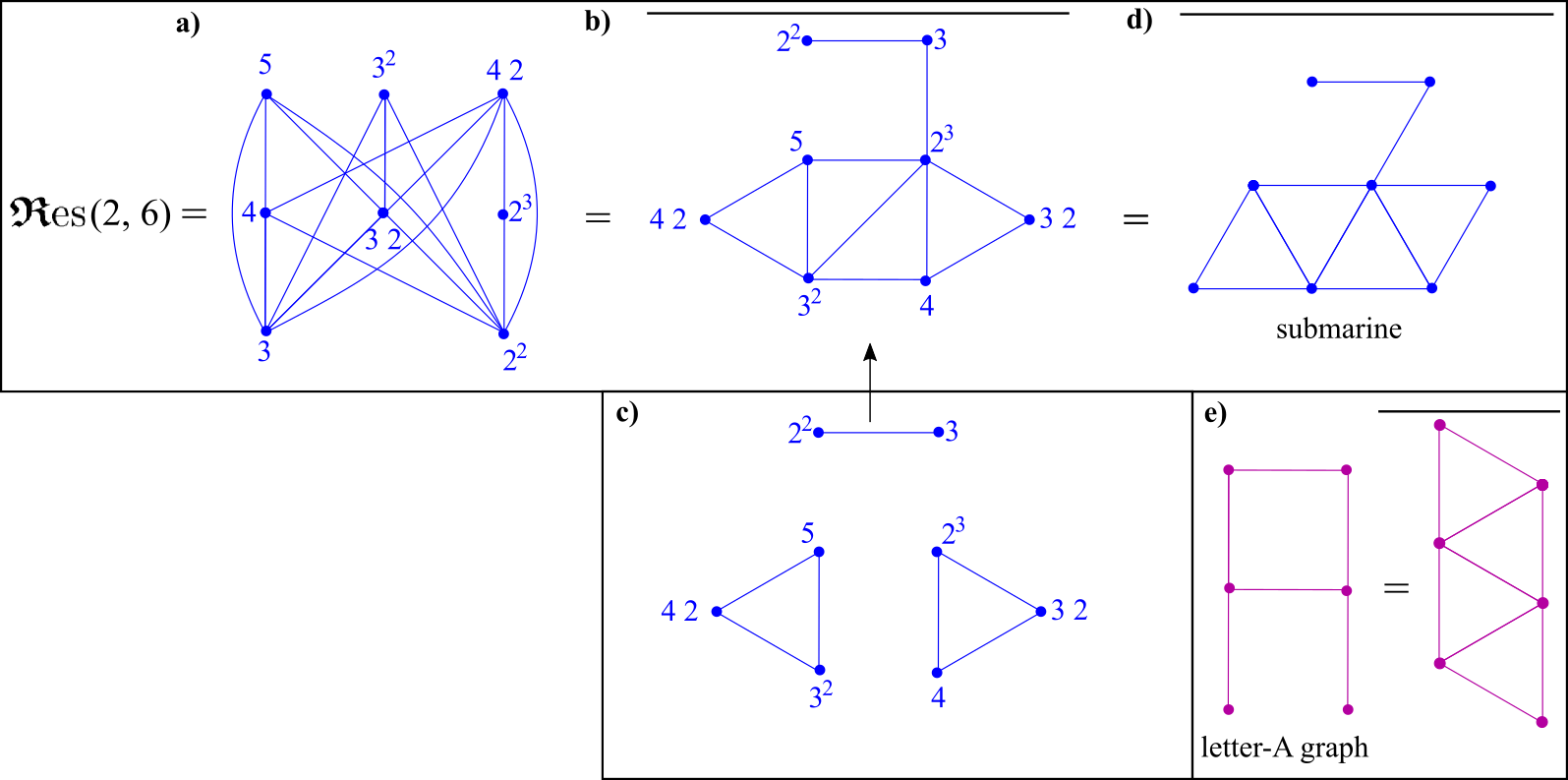}
\caption[Text der im Bilderverzeichnis auftaucht]{        \footnotesize{$\Res(\geq 2, 6)$ in a) tripartite form, 
b) complement form by slight extension of the complete tripartite graph's complement in c), 
d) equal edges and angles `submarine' representation. 
e) Further identification of the submarine's body as the complement of the `letter-A graph' (a small exercise left to the reader). } }
\label{Res-6-2} \end{figure}          }

\m 

\n{\bf Proposition 2} The $(\geq 2, 6)$ residue takes the form of Fig \ref{Res-6-2}.a), with its rather simpler complement in Fig \ref{Res-6-2}.b) and c):  
\be 
\overline{\Res(\geq 2, 6)} 
                       \es  \overline{\mA_4} \, \#_4  \, \mP_2 
                       \=: \mbox{submarine}                                                         \m . 
\ee
\n{\bf Remark 3} This requires a bit of explanation. 

\m 

\n 1) A is the letter-A graph of Fig \ref{Res-6-2}.d).

\m 

\n 2) $\#_4$ means the join at a vertex of valency 4. 
This is a unique prescription in this case since both of $\overline{A}$'s valency-4 vertices are equivalent, 
so which of these is used to make the join makes no difference at the level of unlabelled graphs. 

\m 

\n 3) This `submarine embedding' is not only memorable but also pretty privileged: planar, rectilinear, with all edges of equal length \cite{Atlas} and all angles equal.  
the above join is of the $\mP_2$ `periscope' to the $\overline{\mA}$ `body'.  

\m 

\n{\bf Corollary 2} We can now read off 
\be 
\Top\mbox{-}\Co\mbox{-}\Leib_{\FrS}(\geq 2, 6)   =   \mC_{\sB}(\Res(\geq 2, 6))  
                                                \es  \overline{\mbox{submarine} \, \disjoint \,  \mD_1}           \m , 
\ee  
\be 
\Top\mbox{-}\Leib_{\FrS}(\geq 2, 6)  =  \mC^2_{\sB, \sG}(\Res(\geq 2, 6))
                                    \es  \overline{\mbox{submarine} \, \disjoint \,  \mD_2} 				        \m , 
\ee  
with $\Top\mbox{-}\Co\mbox{-}\Leib_{\FrS}(\geq 2, 6)$ isometric to the previous but with O instead of G as second cone point, and finally  
\be 
\Top\mbox{-}\Leib_{\FrR}(\geq 2, 6)   =   \mC^3_{\sO, \sB, \sG}(\Res(\geq 2, 6))
                                     \es  \overline{\mbox{submarine} \, \disjoint \,  \mD_3}                \m .  
\ee
\n{\bf Remark 4} These are the first nontrivial topological shape graphs according to criteria 0 to 6.  

\m 

\n{\bf Remark 5} $(\geq 2, 6)$'s residue graph is moreover both planar and co-planar, rendering it trivial in senses 8.a) and 8.b).  

\m 

\n{\bf Remark 6} For the general $N$, we have the split-up of Fig \ref{Top-Leib-N-2-Table}. 
The observation that the top row and bottom two rows are just conings generalizes to all $N$, so our concept of residue graph remains useful and is thus justified.   
Note that $\Res$ is ($N - 3$)-partite.   
Once any row attains 5 members, the complement graph contains a $\mK_5$ subgraph and is thus forced to be nonplanar.  
Counting out partitions class by class, it is $N = 8$ which first attains five members in a row. 
This is moreover a bound; it is not precluded that $N = 7$ attains $\overline{\Res}$ planarity by other means; we check this in Sec \ref{Sevens}.  
The formulae for `$\Co$' and `$\Leib$' as cones thereover carry over to all these cases. 
O and G being cone points and thus having equal status follows on from partition refinements forming a lattice with unique top and bottom, subsequently supplemented with 
some extra edges along the lines of Sec \ref{6,2}.
2 is also a cone point, since for the $\geq 2$ case all higher coincidences-or-collisions can be fissioned to leave a binary, 2. 
This ceases to be the case further up the split diagram since say fissioning $2^3$ cannot produce $3$.
%
{            \begin{figure}[!ht]
\centering
\includegraphics[width=0.55\textwidth]{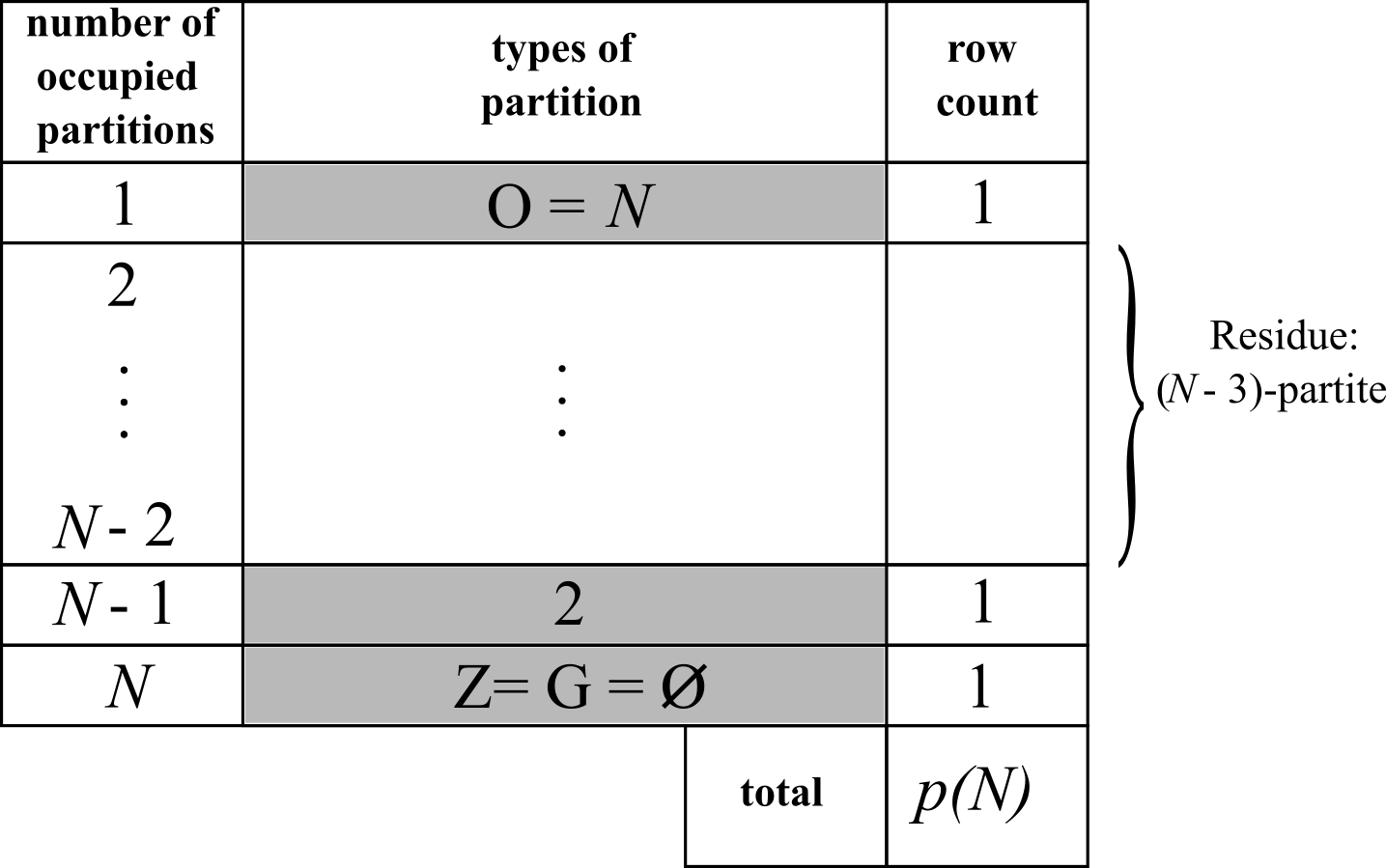}
\caption[Text der im Bilderverzeichnis auftaucht]{        \footnotesize{Schematical form of $(\geq 2, N)$ case's partitions.} }
\label{Top-Leib-N-2-Table} \end{figure}          }

\section{First graphical nontriviality in the ($\mathbb{S}^1$, \biN) class}\label{6,S}

\n{\bf Remark 1} We know the status of this as far as  $N = 5$ by coincidence with the $\geq 2$ case up to that point. 

\m 

\n{\bf Proposition 1} $\overline{\Res}$ is in this case depicted in Fig \ref{Res-6-S} in planar, rectilinear, equal edge lengths and equal edge angles representation. 
We name this embedding of this graph `aircraft carrier' on account of having a $\mK_3$ `control tower' joined to the same $\overline{A}$ `body' in place of the previous `periscope'. 
%
{            \begin{figure}[!ht]
\centering
\includegraphics[width=0.65\textwidth]{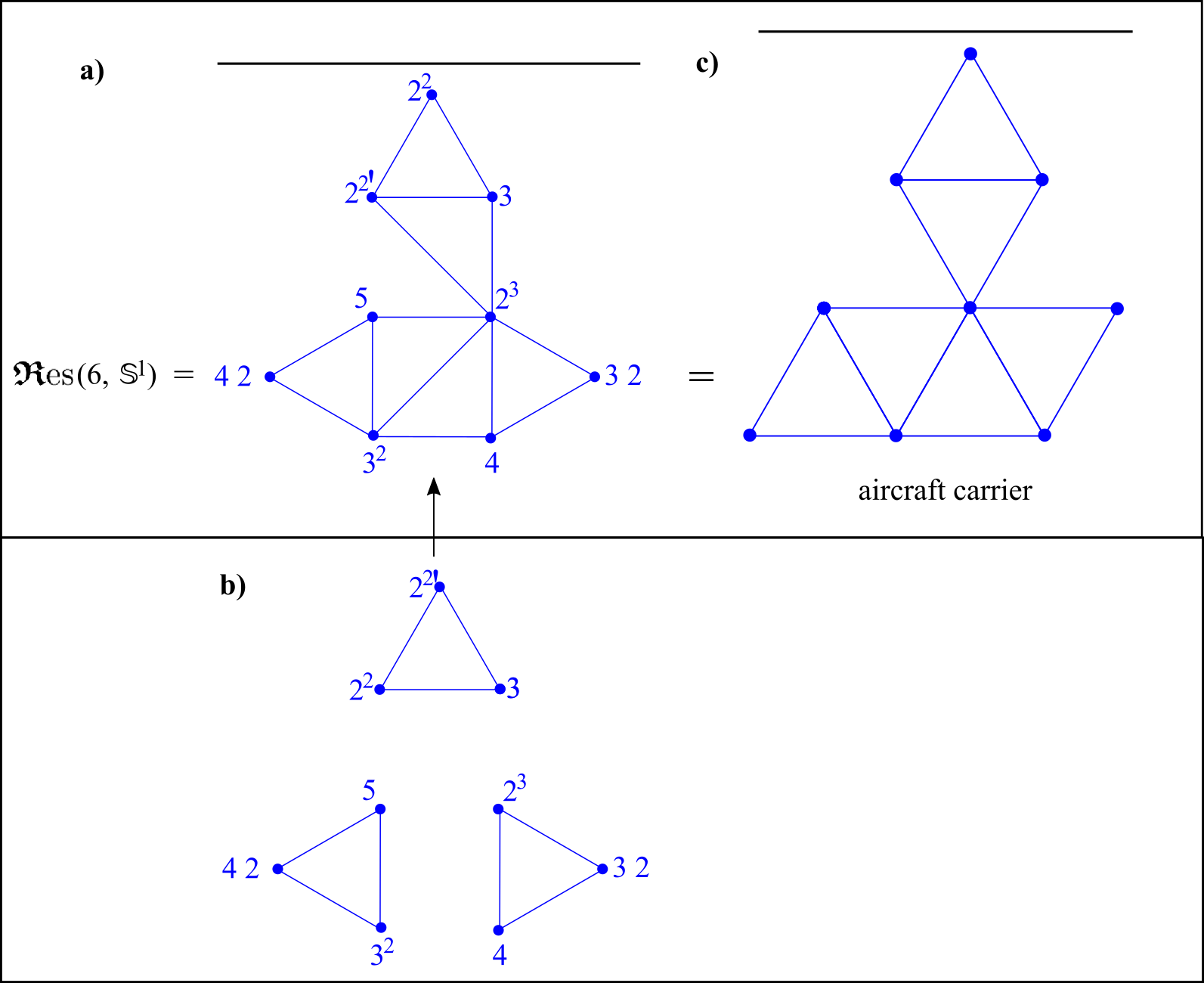}
\caption[Text der im Bilderverzeichnis auftaucht]{        \footnotesize{a) $\Res(\mathbb{S}^1, 6)$ complement, 
from b) a slightly more elaborate extension of this larger complete graph componenents.
c) This also admits an equal edges and angles representation, now in the form of an `aircraft carrier'.   
} }
\label{Res-6-S} \end{figure}          }

\m 

\n{\bf Remark 2} This is nontrivial in senses 0 to 6.
It is planar (triviality 8.a), but its complement is nonplanar, so it is nontrivial in sense 8.b) as well.  
This is because the given figure implies by Corollary B.2. that the original Residue graph is nonplanar.   
In this way, $(\mathbb{S}^1, 6)$ is more complicated than $(\geq 2, 6)$

\m 

\n{\bf Remark 3} The case for general $(\mathbb{S}^1, N)$ is as per Fig \ref{Top-Leib-N-2-Table}, except that $s(N)$ features as the total instead of $p(N)$.   
Qualitative similarities with the previous section abound, but for $N = 6$ upward $s(N) > p(N)$, and the graphs calculated are increasingly different with increasing $N$. 

\m 

\n{\bf Corollary 1} We can now read off 
\be 
\Top\mbox{-}\Co\mbox{-}\Leib_{\sFrR}(\mathbb{S}^1, 6)  =   \mC^2_{\sO, 2}(\Res(\mathbb{S}^1, 6))  
                                     \es  \overline{\mbox{helm} \, \disjoint \,  \mD_2}                       \m , 
\ee  
and 
\be 
\Top\mbox{-}\Leib_{\sFrR}(\mathbb{S}^1, 6)   =   \mC^3_{\sO, \sG, 2}(\Res(\mathbb{S}^1, 6))
                                            \es  \overline{\mbox{helm} \, \disjoint \,  \mD_3}                \m .  
\ee

\section{First graphical nontriviality in the (1, \biN) class}\label{5,1}

\n{\bf Remark 1} We know the status of this as far as  $N = 3$ by coincidence with the $\geq 2$ case up to that point. 

\m

\n{\bf Remark 2} For $N = 4$, the pure-shape case is the gem graph alias 3-fan $\mF_3$. 
While this is nontrivial according to criteria 0 to 5, fans are cones over paths, so this is trivial by criterion 6.
Its residue is moreover the 4-path $\mP_4$ which is clearly [planar], and so trivial in all of criterion 8's ways. 

\m 

{            \begin{figure}[!ht]
\centering
\includegraphics[width=0.55\textwidth]{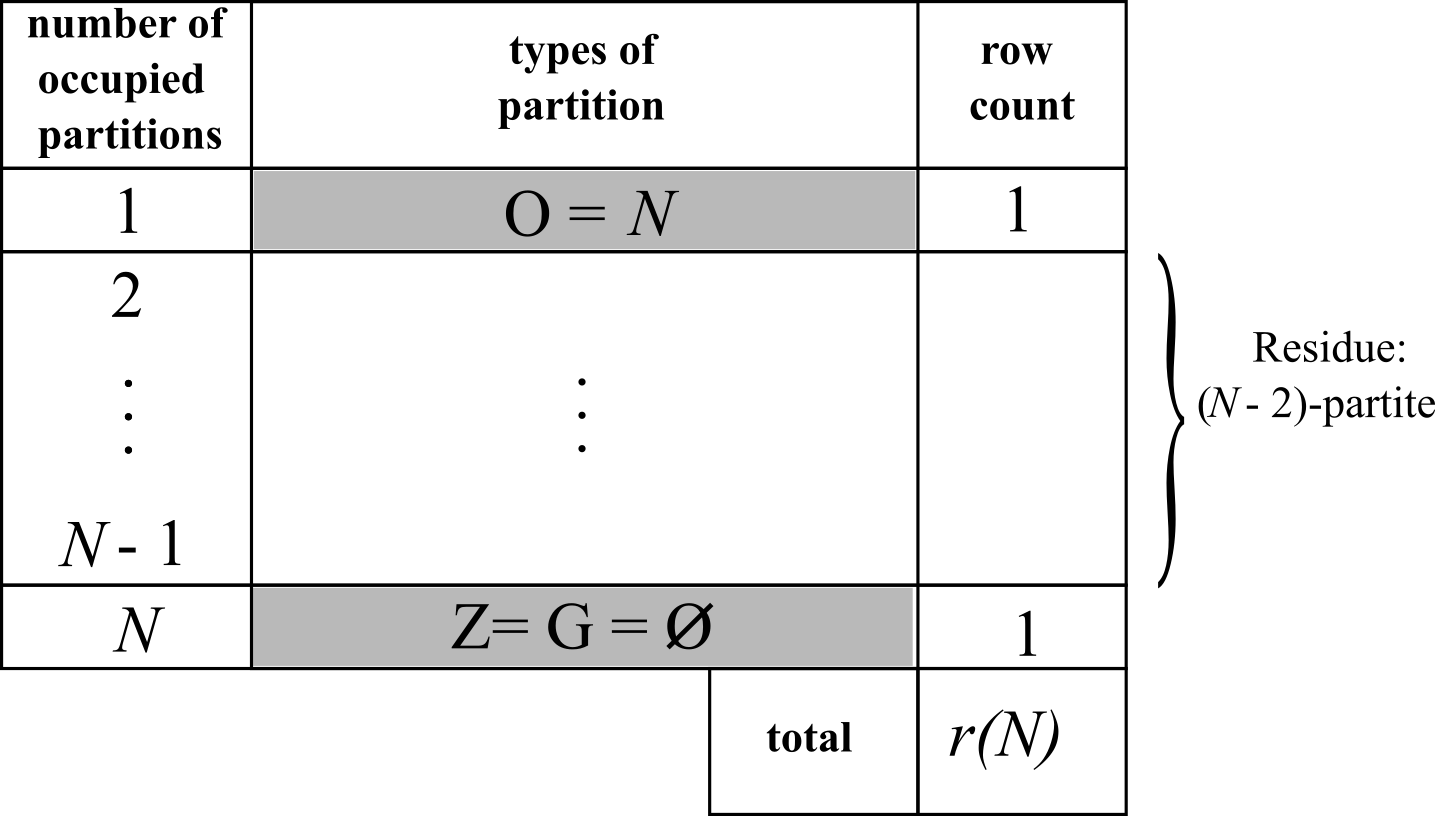}
\caption[Text der im Bilderverzeichnis auftaucht]{        \footnotesize{Schematical form for the $(1, N)$ case.} }
\label{Top-Leib-N-1-Table} \end{figure}          }
%
{            \begin{figure}[!ht]
\centering
\includegraphics[width=0.65\textwidth]{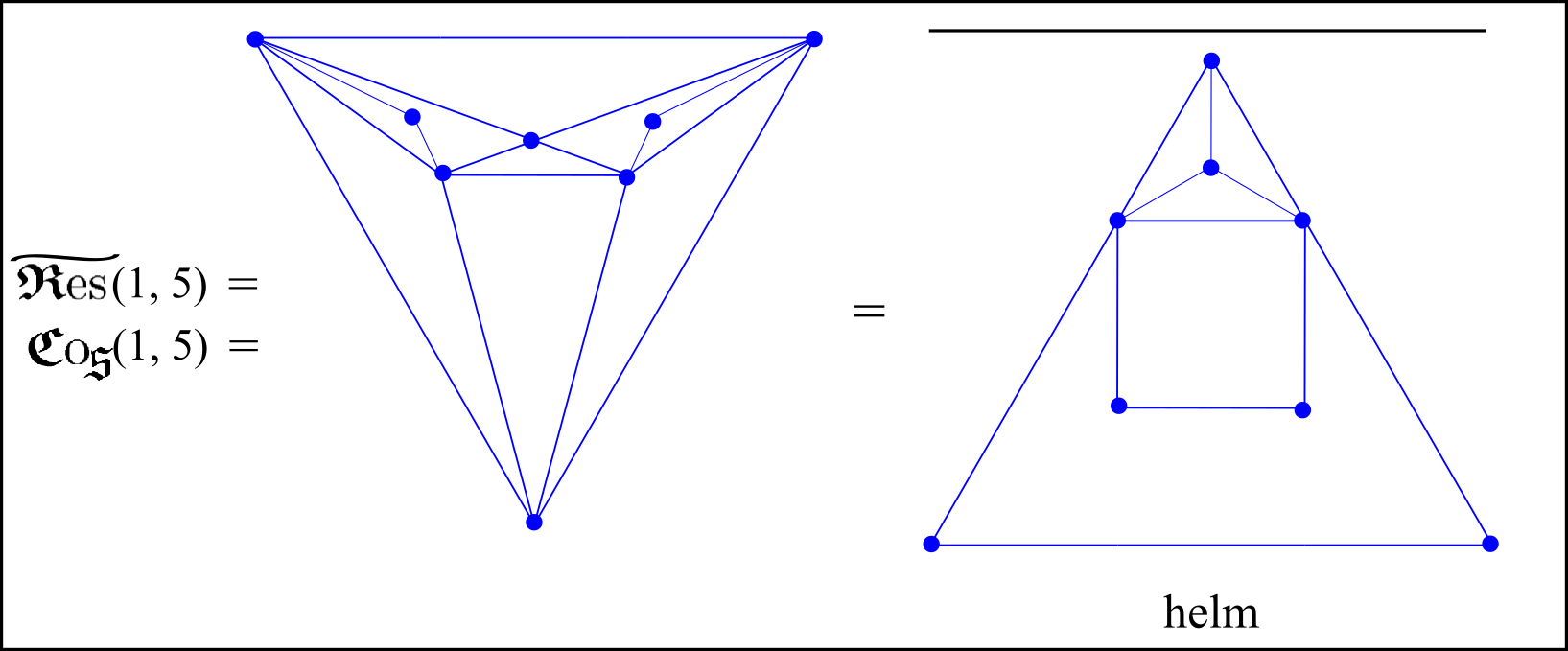}
\caption[Text der im Bilderverzeichnis auftaucht]{        \footnotesize{a) $\w{\Res}(1, 5) = \Co_{\sFrS}(1, 5)$ and b) its complement, which is somewhat simpler (less edges) 
and takes the helm form.} }
\label{Res-5-1} \end{figure}          }

\m 

\n Conceptually, we need a $\widetilde{\Res}$ since the object's subsequent conings have one level less. 
This is however identical with $\Co_{\sFrS}$, so we use that name. 

\m

\n For the 1 class moreover $N = 5$ suffices for graph-theoretic nontriviality, as follows.  
We split up the topological configurations as per Fig \ref{Top-Leib-5-1-Table}.  
Noting the top and bottom rows are just conings (one less than before) 
the remaining content is in the tripartite graph of which Fig \ref{Res-5-1}.a) is a planar representation and Fig \ref{Res-5-1}.b) is a planar representation of the complement.
We refer to this co-graph as `helm', viewing it as a $\mK_4$ crest, noseguard ($\mP_2$ and joins) and wide posterior neck guard (another $\mP_2$ and joins).  

\m 

\n{\bf Proposition 1} 
\be 
\overline{\Res}(1, 5) = \mbox{helm} \m .
\ee 
\n{\bf Corollary 1} We can now read off 
\be 
\Top\mbox{-}\Leib_{\sFrS}(1, 5)  =   \mC_{\sG}(\Res(1, 5))  
                                     \es  \overline{\mbox{helm} \, \disjoint \,  \mD_1}           \m , 
\ee  
with $\Top\mbox{-}\Co\mbox{-}\Leib_{\FrR}(1, 5)$ isometric to the previous but with O instead of G as cone point, and finally    
\be 
\Top\mbox{-}\Leib_{\sFrR}(1, 5)   =   \mC^2_{\sO, \sG}(\Res(1, 5))
                                 \es  \overline{\mbox{helm} \, \disjoint \,  \mD_2}                \m .  
\ee
This satisfies graphical nontriviality conditions 1 to 5 but is however both planar and co-planar.

\section{($\geq$ 2, 7) and ($\mathbb{S}^1$, 7) models}\label{Sevens}
%
{            \begin{figure}[!ht]
\centering
\includegraphics[width=0.45\textwidth]{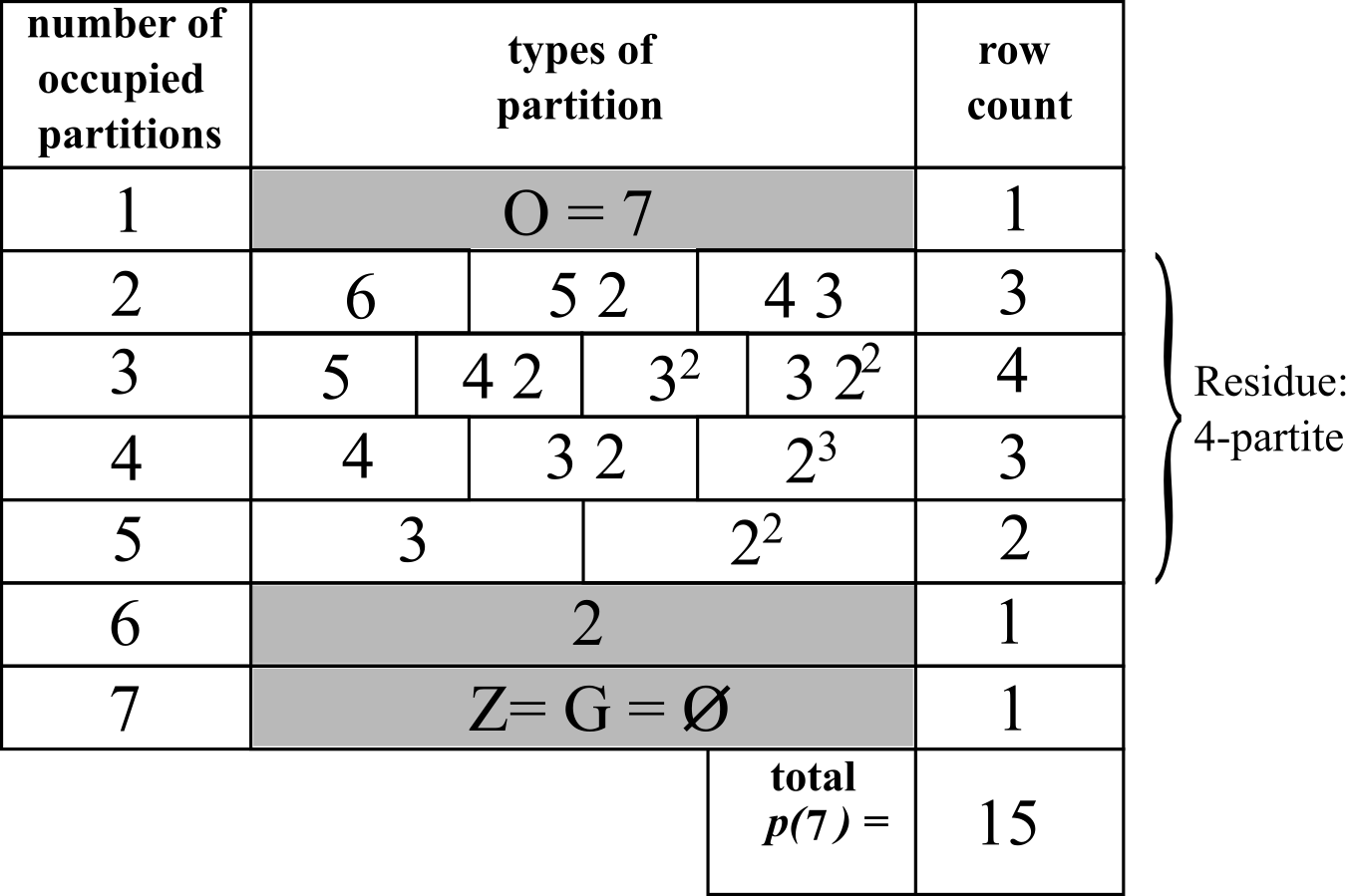}
\caption[Text der im Bilderverzeichnis auftaucht]{        \footnotesize{Partitions of 7 indistinguishable objects, identifying the 15 vertices of $\Res(\geq 2, 7)$.} }
\label{Top-Leib-7-2-Table} \end{figure}          }

\m

\n{\bf Remark 1} For $(\geq 2, 7)$, we have the topological configurations of Fig \ref{Top-Leib-7-2-Table}.

\m 

\n{\bf Remark 2} $\overline{\Res}$ remains manifestly planar, as per Fig \ref{Summary-Table}, but its complement is nonplanar from this Figure and Corollary B.2.    
Thus graphical nontriviality 8.a) is attained but 8.b) is not.  

\m 

\n{\bf Remark 3} We note furthermore that this graph exhibits {\it modularity structure}:  
each $\mK_{p}$ resulting from the 4-partite structure of the split \ref{Top-Leib-7-2-Table} is only linked to `adjacent rungs', 
so the $\mK_p$ can be `strung out in a line' as per Fig \ref{Modul}.
%
{            \begin{figure}[!ht]
\centering
\includegraphics[width=0.6\textwidth]{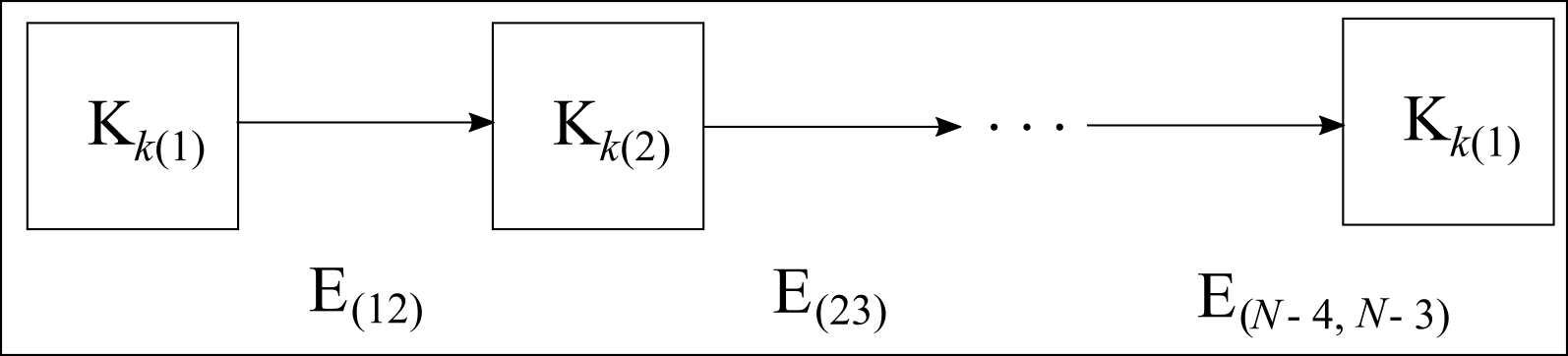}
\caption[Text der im Bilderverzeichnis auftaucht]{        \footnotesize{Schematic form of the current paper's notion of modular graph. 
The edge set E of the graph is partitioned into $\mE_{i, i + 1}$ for $i = 1$ to $N - 3$.
} }
\label{Modul} \end{figure}          }

\m 

\n{\bf Graphical Simplicity Criterion 9} is that the graph is modular:  a finite sequence of $\mK_p$ strung out in a line with further edges only between adjacent $\mK_p$ 
along this line.  

\m 

\n{\bf Remark 4} ($\mathbb{S}^1$, 7)'s topological configurations are as per the further refinement of Fig \ref{Top-Leib-7-2-Table}.  
%
{            \begin{figure}[!ht]
\centering
\includegraphics[width=0.6\textwidth]{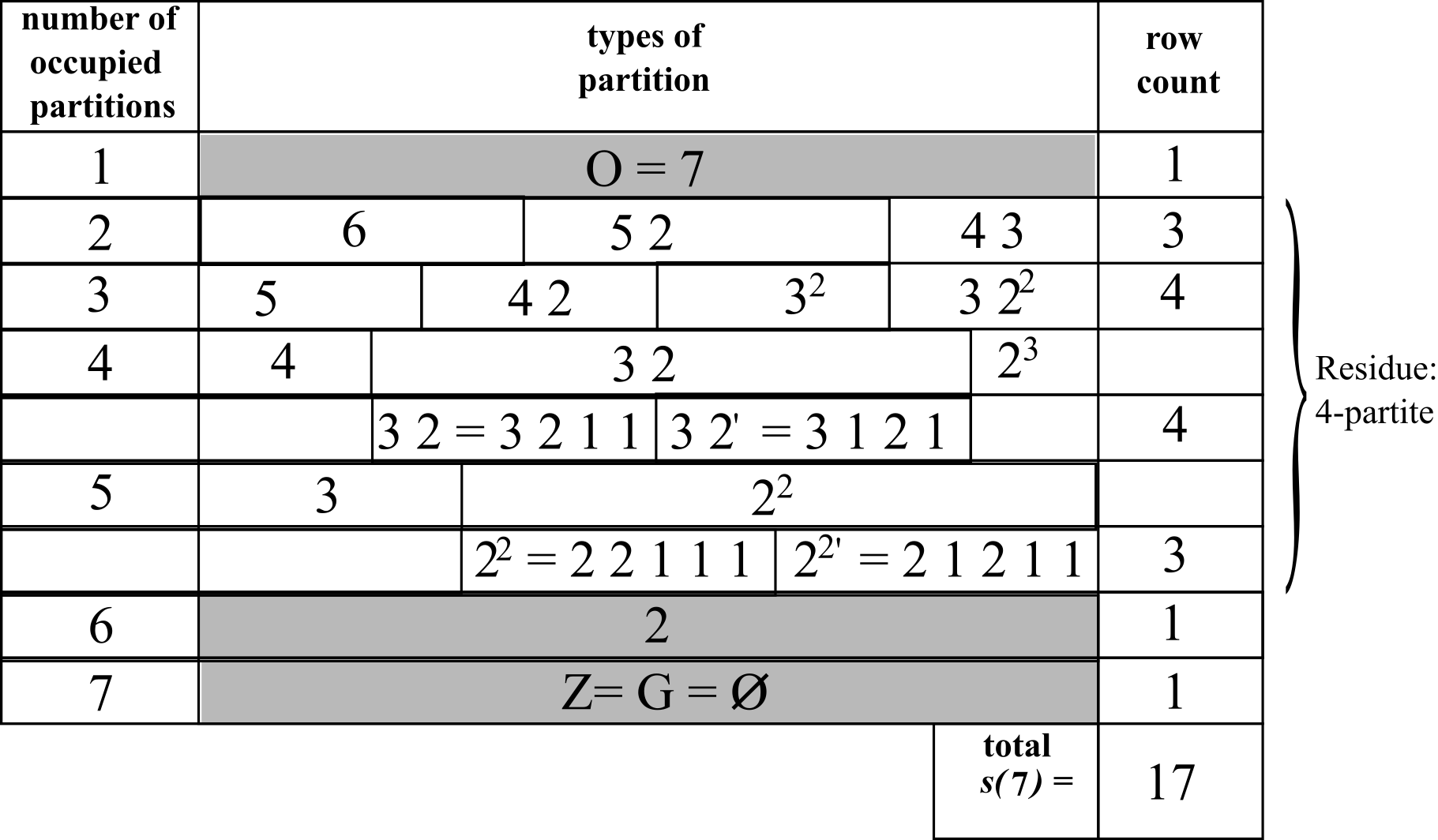}
\caption[Text der im Bilderverzeichnis auftaucht]{        \footnotesize{Identifying the 17 distinct rubber configurations of $(\mathbb{S}^1, 7)$.} }
\label{Top-Leib-7-S-Table} \end{figure}          }

\m 

\n{\bf Remark 5} The corresponding residue graph is likewise modular, as exhibited in Fig \ref{Summary-Table}. 
While this modular presentation has crossings, it can moreover be distorted to exhibit manifest planarity, 
for which a rectilinear representation is given in Fig \ref{Rectilinear}.  
%
{            \begin{figure}[!ht]
\centering
\includegraphics[width=0.5\textwidth]{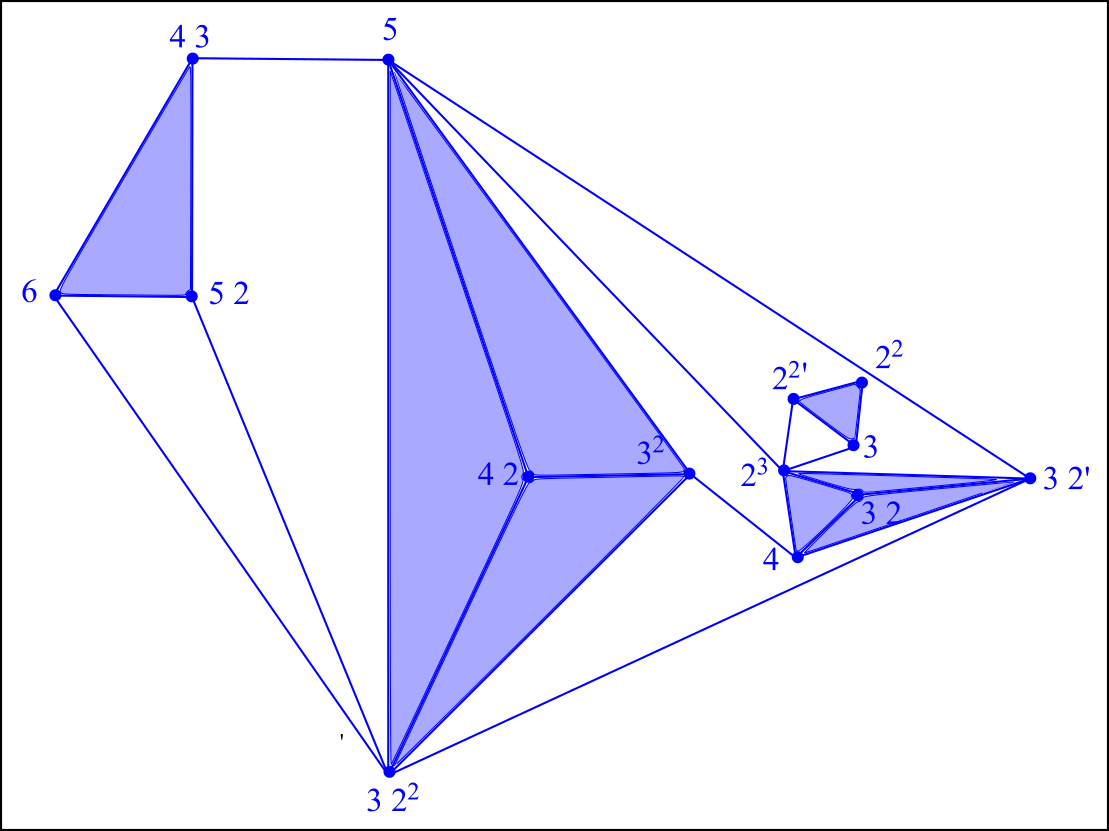}
\caption[Text der im Bilderverzeichnis auftaucht]{        \footnotesize{Rectilinear representation for the $\Res(\mathbb{S}^1, 7)$ graph.
Its modules are shaded blue, for ease of comparison with Fig \ref{Summary-Table}.
} }
\label{Rectilinear} \end{figure}          }

\m 

\n{\bf Remark 6} Thus $(\geq 2, 8)$ and $(\mathbb{S}^1, 8)$ both have the residue planarity and co-planarity of criterion 8.b).  
Thereby, the $N = 8$ bound for nonplanarity is strict for $\geq 2$ and $\mathbb{S}^1$.

\section{(1, 6) model}\label{6,1}

\n{\bf Remark 1} In outline, $\overline{\w{\Res}}(1, 6)$ is both [non-planar] and non-modular, by which it satisfies all of the current article's graphical nontriviality criteria.
Moreover, the combination of being order-18 and non-modular renders it rather too large and complicated to draw; 
we do not moreover need to draw it to demonstrate these nontrivialities. 

\m 

\n It is non-planar because it contains $\mK_5$ subgraphs.  
It is non-co-planar because it contains multiple copies of $\mK_3$ which are not directly linked by edges, so by Corollary B.2, its complement contains a $\mK_{3, 3}$ subgraph. 

\m 

\n It is non-modular because e.g.\ the terminal 2 is adjacent to the central $4^{\prime}$, which is two rungs away.

\section{Larger \biN}\label{Big-N}

\n{\bf Remark 1} Increase in complexity of coincidence-or-collision structure is considerable with increasing $N$. 
The growth of $p(N)$ with $N$ in the Hardy--Ramanujan asymptotic formula (\ref{HR}) gives a rough indication of the problem. 
This formula of course serves just as well for the $\geq 2$ case's residue graphs' $p(N) - 3$.  
This points to topological shapes' coincidence-or-collision structure becomes a major issue in Shape Theory and $N$-body problem even for just physically or statistically 
modest values of $N$ such as 30, 100, 300, 1000 or 3000, for which the estimates for $p(N)$ are around $6000$, $10^8$, $10^{16}$, $10^{31}$ and $10^{56}$ respectively.   

\m 

\n{\bf Remark 2} It is furtherly pertinent however to estimate the number of edges involved. 
A simple estimate for this follows from the Hardy--Ramanujan asymptotic formula (\ref{HR}) and the semi-saturation bound (\ref{Semi-Sat}) on edge number for [graphs]:  
\be 
e(N)  \m \approx \m  \frac{1}{192 \,  \, N^2} \, \mbox{exp}
\left(2 \, \pi \sqrt{\frac{2 \, N}{3}    } 
\right)                                                    \m 
\mbox{ as } \m  N \longrightarrow \infty                   \m .  
\ee
For $N = 30$, 100, 300, 1000 and 3000, this returns $10^7$, $10^{15}$, $10^{31}$, $10^{62}$ and $10^{111}$.
This estimate could furthermore be improved by obtaining a residue-complement-graph-specific bound on unsaturation.  

\m 

\n{\bf Remark 3} While $p(N)$ provides a lower bound on $s(N)$ and $r(N)$, it would be useful to obtain asymptotic formulae for $s(N)$ and $r(N)$ themselves.

\vspace{10in}

\section{Conclusion}\label{Conclusion}

\n We considered a rubber alias topological notion of shapes (or scaled shapes). 
This gives a less structured Shape(-and-Scale) Theory (subsequently collectively referred to as Relational Theory) than the previous geometrical (scaled) shape versions 
considered along the lines of Kendall's work. 
Our less structured version is simpler in various ways while none the less encoding some features of geometrical (scaled) shape theories.
One way in which it is simpler is that there are only three topological (scaled) shape theories for connected manifold without boundary carrier spaces $\FrC^d$. 
In particular, for carrier space dimension $d \geq 2$, there is just the one topological (scaled) shape theory independent of any further detail of $d$ 
or of any further topological detail of $\FrC^d$. 
For $d$ = 1, on the other hand, distinction is maintained between $\mathbb{R}$ and $\mathbb{S}^1$ carrier spaces.  

\m 

\n The $d \geq 2$ topological scaled shape theory's configurations are in 1 : 1 correspondence with the well-studied partitions, facilitating pursuit of this case to some extent. 
In the pure-shape counterpart, the maximal coincidence-or-collision (corresponding to the single-part partition) alone is excluded.  
The topologicial adjacency condition between such configurations -- which retains meaning at the level of all the corresponding geometrical (scaled) shape theories, 
furthermore equips the space of rubber (scaled) shapes as graphs. 
This is a second simplyfying feature of rubber rather than geometrical (scaled) shapes, since the latter's configuration spaces are in general stratified manifolds 
-- much more complicated and less familiar objects than the rubber relational spaces' graphs. 
These graphs moreover in general have more edges than those in the partition refinement lattice (which, in undirected form, features therein as a subgraph).  
On the other hand, the two $d = 1$ topological scaled shape theories' configurations further refine the partitions in two stages, with the $\mathbb{S}^1$ theory 
providing a first refinement and the $\mathbb{R}$ theory providing a subsequent refinement.  
$\mathbb{R}$ refinement first occurs for $N = 4$ \cite{AF, FileR, II}, whereas $\mathbb{S}^1$ refinement first occurs for $N = 6$.  
%
{            \begin{figure}[!ht]
\centering
\includegraphics[width=0.35\textwidth]{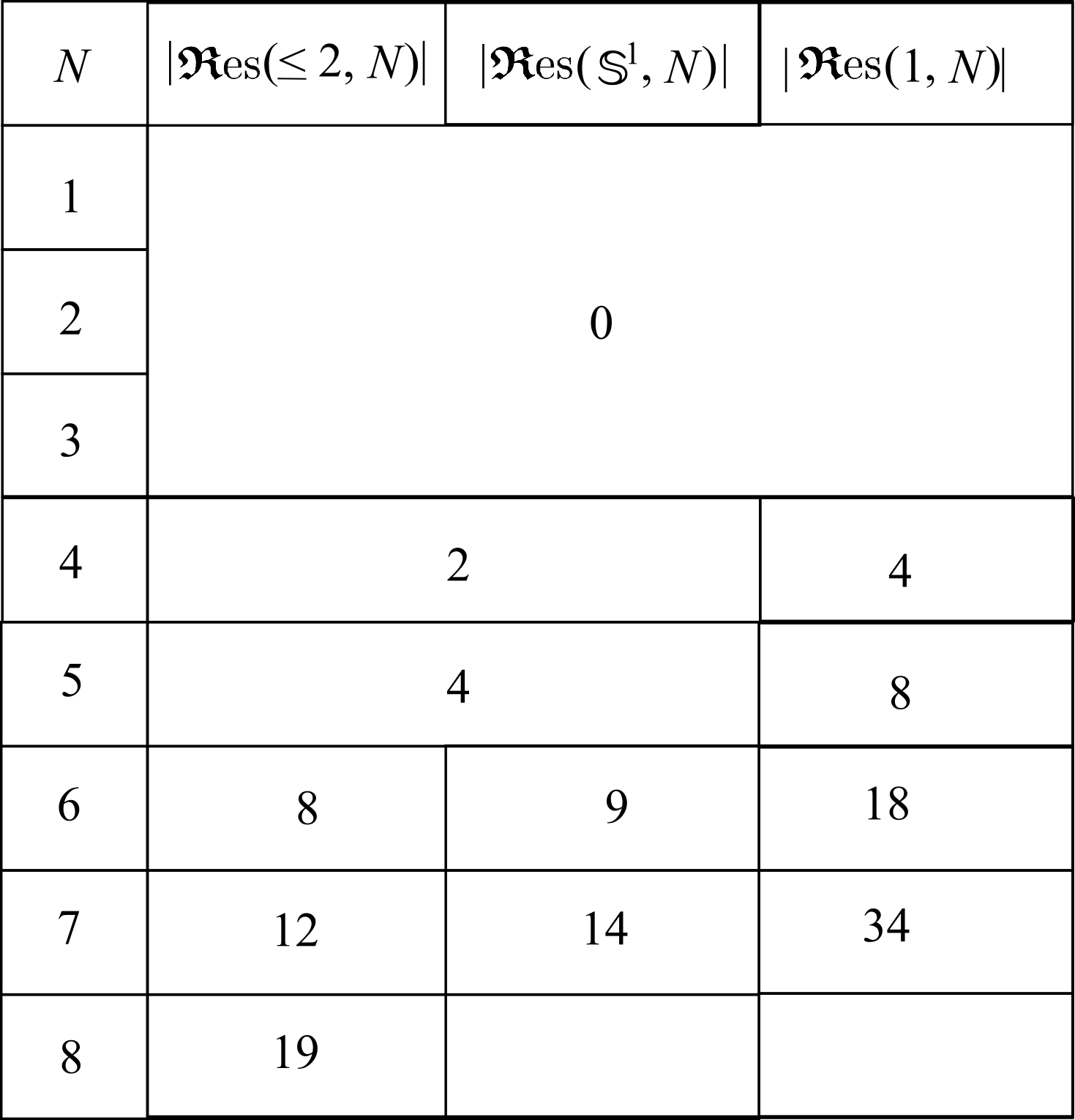}
\caption[Text der im Bilderverzeichnis auftaucht]{        \footnotesize{$(\tau, N)$ grid of orders of topological Leibniz space residue graphs' dimensions.} }
\label{Res-Graph-Order} \end{figure}          }

\m 

\n We looked in particular at which topological Leibniz space graphs meet some nontriviality criteria. 
A preliminary such criterion is distinction from the partitions, for which (1, 4) and $(\mathbb{S}^1, 6)$ are minimal by the previous paragraph. 
Our next approach is to decone the topological Leibniz space graphs so as to remove $N$-independent features (bar for the very smallest $N$, for which conflations occur). 
Namely that the generic configuration, and the maximal coincidence-or-collision when present: in the scaled case, are topologically adjacent to everything and so are cone points. 
Furthermore, for $d \geq 2$ and $\mathbb{S}^1$, the binary coincidence-or-collision is as well, but this ceases to old for the $\mathbb{R}$ carrier space from $N = 4$ upward.
Removing these cone points leaves us with a nontrivial residue piece, $\Res$, the orders of which graphs are sumarized in Fig \ref{Res-Graph-Order}. 
This distills Fig \ref{Top-Space-Graph-Order}'s counts into a single scale-independent count excluding the cone points, 
which are universally-present and thus {\sl undistinctive} features.  
Furthermore, as Fig \ref{Sat-Summary-Table} indicates, the complements of these have less edges (bar for the very smallest $N$ again), and so are easier to characterize and depict. 
Thus we concentrate on topological Leibniz space residue complements in our summary table figure \ref{Summary-Table}.  
%
{            \begin{figure}[!ht]
\centering
\includegraphics[width=0.35\textwidth]{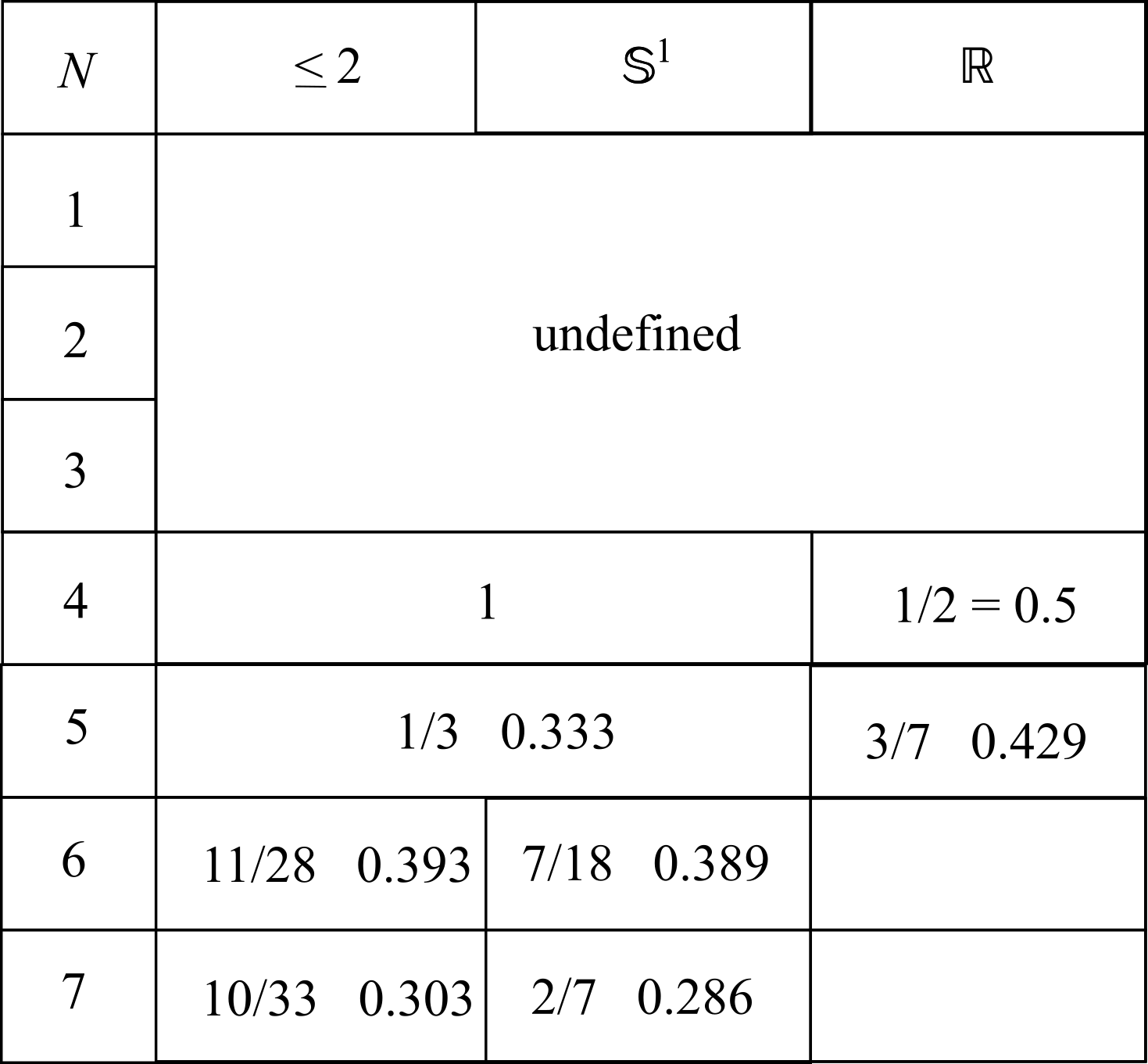}
\caption[Text der im Bilderverzeichnis auftaucht]{        \footnotesize{Saturation of Leibniz space residue complements, 
indicating these to be simpler than the residues themselves.} }
\label{Sat-Summary-Table} \end{figure}          }

\m

\n A first set of non-triviality criteria is that the residue complement be more than jut a collection of points, paths or cycles. 
This picks out $(\geq 2, 6)$, $(\mathbb{S}^1, 6)$ and (1, 5). 
While all three of these are planar, $(\mathbb{S}^1, 6)$ moreover has a nonplanar complement. 
Nonplanarity of both the residue graph and its complement first occur for (1, 6), $(\geq 2, 8)$ and $(\mathbb{S}^1, 8)$: a second non-triviality criterion. 
This moreover coincides with the realization of a third non-triviality condition: the breakdown of modularity. 
This is in the sense of the graph being a ladder of complete subgraph $\mK_n$ blocks each only attached to adjacent blocks along the ladder. 
N.B.\ that our summary Figure is modularly presented (at the expense of manifestly exhibiting planarity in one graph: the $(\mathbb{S}^1, 7)$ residue complement). 
%
{            \begin{figure}[!ht]
\centering
\includegraphics[width=1.0\textwidth]{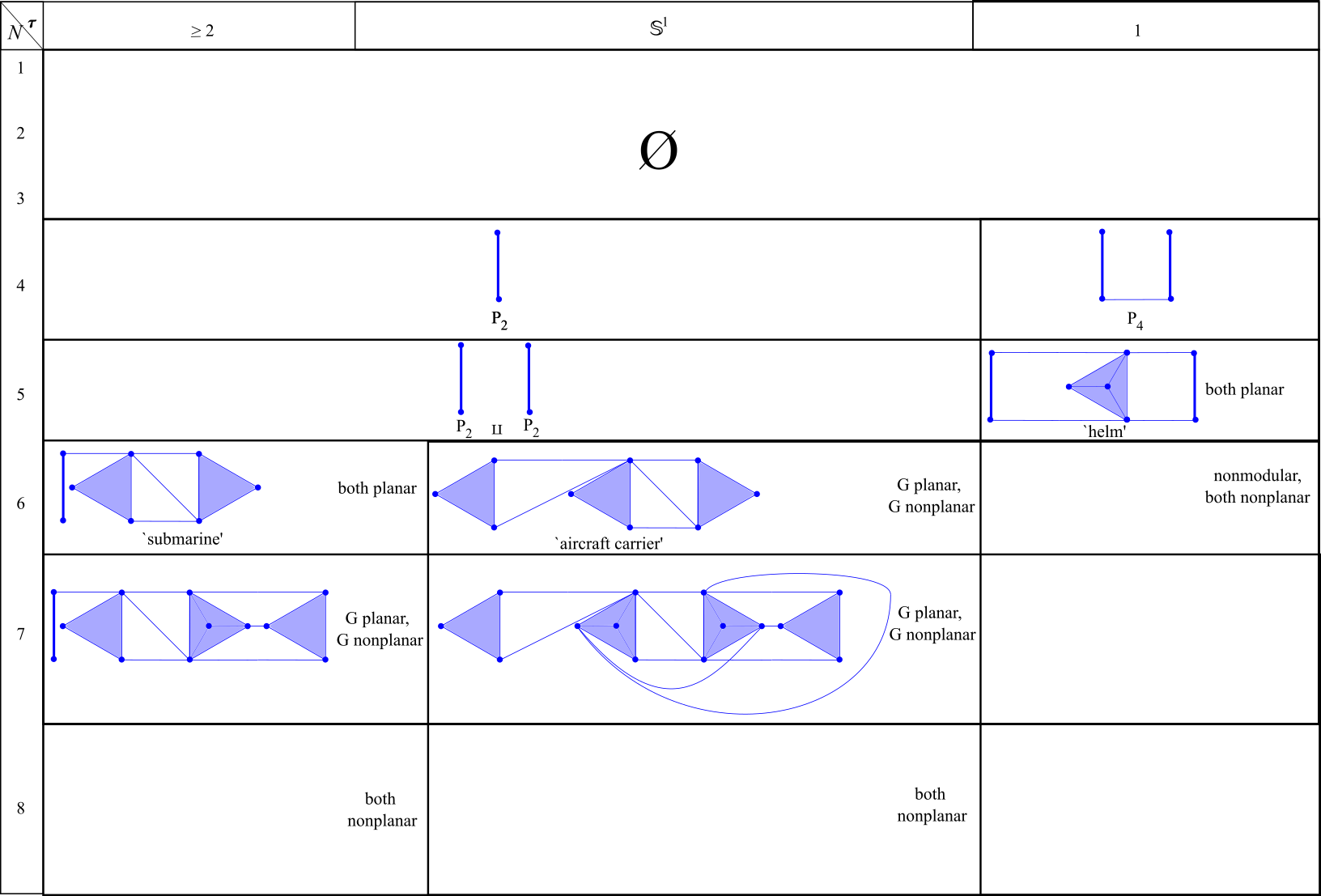}
\caption[Text der im Bilderverzeichnis auftaucht]{        \footnotesize{Summary table of topological Leibniz space residue complements.} }
\label{Summary-Table} \end{figure}          }

\m 

\n{\bf Further frontiers. 1. \biN-body problem} 

\m 

\n Metric-level scaled shape theory's orbit space is known \cite{ML00} to attain a genericity criterion for $N \geq 5$ in 3-$d$, 
at the level of the full diversity of orbits and the corresponding stabilizers (alias isotropy groups) being realized.  
On the other hand, the current article's less structured rubber (scaled) shapes analysis has revealed nontrivialities for which $N = 6$ and $N = 8$ are minimal.  
This points to qualitative limitations in considering $N$-body problems only up to $N = 5$. 
For further interplay between these results and yet further sources of qualitative effects for the $N \geq 6$ body problem, see \cite{Minimal-N}.

\m 

\n{\bf Further frontiers. 2. Model arena of GR with topology change} 

\m 

\n The recent article \cite{ACirc} showed that relational theories formulated via an absolute space with subsequent quotienting out of automorphisms 
none the less remember features of the absolute space thus `indirectly involved', at the topological level, even in the 1-$d$ case of $\mathbb{R}$ versus $\mathbb{S}^1$.  

\m 

\n One longstanding approach elsewhere in the Theoretical Physics literature is to 
eliminate topological dependence by summing over `all topologies' \cite{Mis-Top, Witten89, GH92} (maybe within some class).
As \cite{ACirc} also outlined, this can also be applied in the formulation of topologically background independent theories. 

\m 

\n At the {\sl geometrical} level, moreover, sums over `all topologies' rapidly get out of hand, in particular in the most physically and mathematically interesting 
dimensions 3 and 4.  
The current subsection serves to point out that at the {\sl rubber} level, sums of `all topologies' collapse to summing over {\sl just three} universality classes: 
$\mathbb{R}$, $\mathbb{S}^1$ and $\FrC^d$ for $d \geq 2$ (for the connected manifold without boundary class of absolute spaces).   
This is a far more tractable model; indeed, one can argue that the physically interesting dimensions do not include 1 anyway, 
in which case the $\FrC^d$ class's universality gives all.  
There is a further caveat that scale-invariant absolute spaces admit a scaled and pure-shape pair of relational theories, 
whereas scale-dependent ones render the first of these obligatory. 
This doubling affects just the $d \geq 2$ class, by which summing over 4 and 2 topologies rather than 3 or 1 is necessary for some purposes. 
But in all instances, these are finite sums and with very few terms indeed.  

\m 

\n The General Relativity structures to compare this to are Fischer's Big Superspace and its further generalization \cite{Top-Shapes} to Grand Superspace, as outlined below. 

\m 

\n All of the models in question involve a configuration space of the schematic form 
\be
\Big\mbox{-}\FrQ = \coprod_{\tau \in \sFrT} \FrQ(\tau)  \m .  
\ee
Such configuration spaces might additionally be accorded further levels of mathematical structure, such as their own topology.  

\m 

\n At the level of actions, quantum operators, quantum path integrals, notions of information...  
this subsequently involves using corresponding `sums over topologies' to remove the effect of the element of choice of a particular topology.
Schematically, one replaces $\tau$-dependent objects $\FrO(\tau)$ by 
\be
\mbox{\Large S}_{\tau \in \sFrT} \FrO(\tau)                           \m .
\ee
\n{\bf Example 1} Let $\FrT$ be the set of all topological manifolds $\FrT$ (rather than topological spaces more generally, and taken to exclude cases `with boundary' and so on).  
Further common class restrictions include the following. 

\m  

\n 1) Considering only the connected manifolds \cite{GH92}.  

\m 

\n 2) Considering only the compact manifolds \cite{Battelle}.

\m 

\n 3) Considering only the orientable manifolds.

\m 

\n 4) Considering only topological manifolds of a fixed dimension.

\m 

\n In the context of point-or-particle models, variable $N$ can also be considered to have topological content \cite{ABook}
(at the level of allowing point-or-particle fission and fusion and/or particle creation and annihilation).
Because of this, 
\be
\coprod_{N \in \mathbb{N}_0} \m \mbox{ and } \m \sum_{N \in \mathbb{N}_0} 
\ee
operations are also an option among the various features to be considered.  

\m 

\n One possibility is then 
\be
\Big\mbox{-}\Top\mbox{-}\FrR(d)  \es  \coprod_{N \in \mathbb{N}_0} \, \Top\mbox{-}\FrR(\FrC^d, N) 
\label{Big-2-T}
\ee 
for any $d \geq 2$, or 
\be 
\Big\mbox{-}\Top\mbox{-}\FrR(1)  \es \coprod_{\sFrC^1 \in \mathbb{R}, \, \mathbb{S}^1 } \,\coprod_{N \in \mathbb{N}_0} \, \Top\mbox{-}\FrR(\FrC^1, N) \m  
\label{Big-1-T}
\ee 
for $d = 1$.   

\m 

\n One finally has 
\be 
\Grand\mbox{-}\FrR  \es  \coprod_{d = 1}^{\geq 2} \,  \coprod_{\sFrC^d \in \sFrM(\mbox{\scriptsize connected)}} \, \coprod_{N \in \mathbb{N}_0} \, \FrR(\FrC^d, N)  
                    \es \coprod_{\sFrC^d = \mathbb{R}, \, \mathbb{S}^1, d \geq 2} \, \coprod_{N \in \mathbb{N}_0} \, \FrR(\FrC^d, N) \m , 
\label{Grand-T}
\ee 
which is the disjoint union of 3 (or 4 if scaled and pure-shape $\d \geq 2$ are both relevant) graphs per $N$.  

\m 

\n GR counterparts of point-or-particle relational spaces, on the other hand, take the following often-mentioned (if in practise only formally treated) form. 

\m 

\n 0) for a given 3-$d$ spatial topology $\bupSigma_3$, 
\be 
\Riem(\bupSigma_3)
\ee 
is the space of all positive-definite 3-metrics on $\bupSigma_3$.  
This is GR's analogue of constellation space. 

\m 

\n 1) The group of physically irrelevant automorphisms usually considered in this case are $Diff(\bupSigma_3)$: the 3-diffeomorphisms of the corresponding $\bupSigma_3$.  

\m 

\n 2) The corresponding quotient is Wheeler's \cite{Battelle} 
\be 
\Superspace(\bupSigma_3) \es \frac{\Riem(\bupSigma_3)}{Diff(\bupSigma_3)}                                                 \m ,
\ee  
further studied in \cite{DeWitt67, DeWitt, Fischer70, FM96, Giu09}. 
A particular feature that Superspace shares with relational spaces is that it is stratified; it is now 3-metrics possessing Killing vectors that constitute the nontrivial strata. 

\m 

\n 3) Fischer \cite{Fischer70} moreover entertained the further concept of a 
\be 
\Big\mbox{-}\Superspace(3) \es \coprod_{\bupSigma_3 \in \sFrT^3} \, \Superspace(\bupSigma_3)                    \m . 
\ee 
This would usually be considered for $\sFrT^3$ additionally connected, compact and orientable.
It is a model underlying a variant of GR which furthermore allows for spatial topology change; by technical necessity, it remains a merely formal model.  

\m 

\n 4) With $d$-dimensional superspace being as straightforward to define, one can additionally conceive of 
\be 
\Grand\mbox{-}\Superspace  \es \coprod_{d = 1}^\infty \, \coprod_{\bupSigma_d \in \sFrT^d} \, \Superspace(\bupSigma_d)   \m ,  
\ee 
possibly suppressing $d = 1$ and 2 contributions since these have no degrees of freedom for GR. 
This is now for a variant of GR which allows for topology change including change of spatial dimension.  

\m 

\n In this light, one can view (\ref{Big-1-T}) and (\ref{Big-2-T}) (with or without variable-$N$) as models of Fischer's Big Superspace, 
whereas (\ref{Grand-T}) provides a calculable analogue even of Grand Superspace (for $N$ fixed and of moderate size).  

\m 

\n{\bf Further frontiers. 3. Yet more general carrier spaces} 

\m 

\n One would here consider dropping between some and all of the connected, manifold and without boundary specifications of the carrier spaces used so far 
in the theory of topological (and for that matter geometrical) shapes and scaled shapes.  

\m 

\n{\bf Acknowledgments} I thank Chris Isham and Don Page for discussions about configuration space topology, geometry, quantization and background independence. 
I also thank Jeremy Butterfield and Christopher Small for encouragement. 
I thank Don, Jeremy, Enrique Alvarez, Reza Tavakol and Malcolm MacCallum for support with my career. 
This paper is dedicated to all those who dare to think interdisciplinarily.

\vspace{10in}
  
\begin{appendices}

\section{Supporting Partition Theory}\label{App-P}

\n{\bf Definition 1} A {\it partition} of a natural number $N$ is an split of it into unordered positive-integer summands, called the {\it parts} of that partition.
One can think of this as splitting up a collection of items between boxes.

\m 

\n{\bf Remark 1} A simple notation for this is decomposition into summands, e.g.\ 3 + 1 + 1 is a partition of $N = 5$. 
A more efficient notation for this is $3 \, 1^2$. 
In the context of a particular fixed $N$, one can furthermore remove reference to the single-ocupancy boxes, by which this example's partition is denoted just by 3.  

\m 

\n{\bf Definition 2} The {\it partition number} $p(N)$ \cite{Combi, Conway, Hardy-Wright} is the number of ways that $N$ can be partitioned.

\m 

\n{\bf Remark 2} Evaluating $p(N)$ is an interesting and nontrivial problem, for which considerable resources exist \cite{p(N)}. 
For the current article, we need $p(1)$ to $p(8)$, which are as per Fig \ref{PN-Table}.
%
{            \begin{figure}[!ht]
\centering
\includegraphics[width=0.27\textwidth]{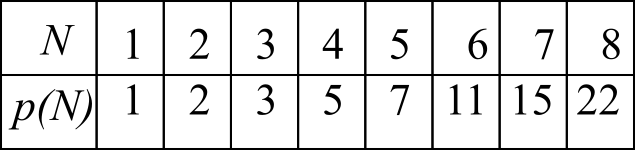}
\caption[Text der im Bilderverzeichnis auftaucht]{        \footnotesize{$p(N)$ for the current article's range of small $N$.} }
\label{PN-Table} \end{figure}          }

\m

\n The current article furthermore both splits partitions for a fixed $N$ by the number $M$ of boxes occupied, $q(N, M)$ and considers the partition refinement lattice (Appendix C). 

\m  

\n{\bf Remark 3} The current article's final use of partition theory involves the \cite{HR18}  
\be 
\mbox{\it Hardy--Ramanujan asymptotic formula }            \m 
p(N)  \m \sim \m  \frac{1}{4\sqrt{3} \, N} \, \mbox{exp}
\left(\pi \sqrt{    \frac{2 \, N}{3}    } 
\right)                                                    \m 
\mbox{ as } \m  N \longrightarrow \infty                   \m .
\label{HR}
\ee 
\n{\bf Remark 4} Whether the current article's real-line and circular topological scaled shape counts $r(N)$ and $s(N)$ occur elsewhere in mathematics is left for now as an open question.

\section{Supporting Graph Theory}\label{App-G}

\subsection{Introductory concepts}\label{G-1}

\n{\bf Definition 1} A {\it graph} $\mG$ consists of a set of {\it vertices} $V$ and a set of {\it edges} $E$ interlinking some subset of the vertices.  

\m 

\n{\bf Remark 1} The graphs used in the current article are all finite. 
They are `simple graphs' rather than `multigraphs' in the senses of having, 
firstly, at most one edge between any two given vertices, and, 
secondly, of having no loops (edges running between a vertex and itself).
The current article's graphs are furthermore {\it labelled}, in the sense that their vertices are decorated with labels (partitions or refinements thereof).  

\m 

\n{\bf Definition 2} A graph $\mG$'s {\it order} is its number of vertices, denoted by $|\mG|$, whereas its {\it size} e(G) is its number of edges.  
The {\it degree} alias {\it valency} $\md(v)$ of a vertex $v$ is the number of edges emanating from it.

\m 

\n{\bf Definition 3} A {\it path graph} $\mP_n$ is an alternating sequence 
\be 
v_0 e_1 v_1... e_{n - 1} v_n   \m ,
\ee  
where the vertex set is $V = \{v_i: i = 1 \mbox{ to } n\}$ 
and   the edge   set is $E = \{e_j: j = 1 \mbox{ to } n - 1\}$.  
So these are connected graphs in which all but two of the valencies $\md(v) = 2$; 
the exceptions to this are the end-point vertices $v_0$ and $v_n$, of valency $\md(v) = 1$. 

\m 

\n{\bf Definition 4} A graph $\mG$ is {\it connected} if it contains a path between any two of its vertices.
$\mG$ is {\it disconnected} if it is not connected.
The {\it components} $\mQ_i$, $i = 1$ to $k$ of $\mG$ are its maximal connected subsets; these partition $\mG$. 
We denote graphs with multiple components by the disjoint union of the names of their components,
\be 
\mG = \coprod_{i = 1}^k \mQ_i     \m .  
\ee  
\n{\bf Remark 2} If multiple copies of a given component $\mH$ are present in such a disjoint union, we use the power notation 
\be 
\mH^n = \, \coprod_{i = 1}^n \mH  \m .
\ee  

\m 

\n{\bf Definition 5} A {\it totally disconnected graph} is one with no edges at all.
We denote these by $\mD_n$, where $n = |\mG|$. 
Less than Graph Theory suffices to consider these, since by having no edge structure, 
the definition of graph collapses in this case to just the definition of a point set. 

\m 

\n{\bf Remark 3} Less than Graph Theory also suffices for paths, with now ordered point sets sufficing to describe these. 

\m 

\n{\bf Definition 6} A {\it cycle graph} $\mC_n$ is an alternating sequence as above except that $v_n$ is now $v_0$ again: 
\be 
v_0 e_1 v_1... e_{n - 1} v_0   \m .
\ee   
So these are connected graphs in which all the valencies are 2.  
This is because a cyclic `joining of the dots' order, alias an order modulo periodicity and choice of starting point, suffices to describe these. 

\m 

\n{\bf Definition 7} A {\it complete graph} $\mK_n$ is one in which there is an edge between any two vertices. 
So all vertex valencies are $|\mG| - 1 = n - 1$. 

\m 

\n{\bf Remark 4} See Fig \ref{Graph-1} for the first few members of each of the above series. 
This includes those examples which are small enough to be multiple of the above, and subsequently the first distinct members of each of these series. 
%
{            \begin{figure}[!ht]
\centering
\includegraphics[width=0.9\textwidth]{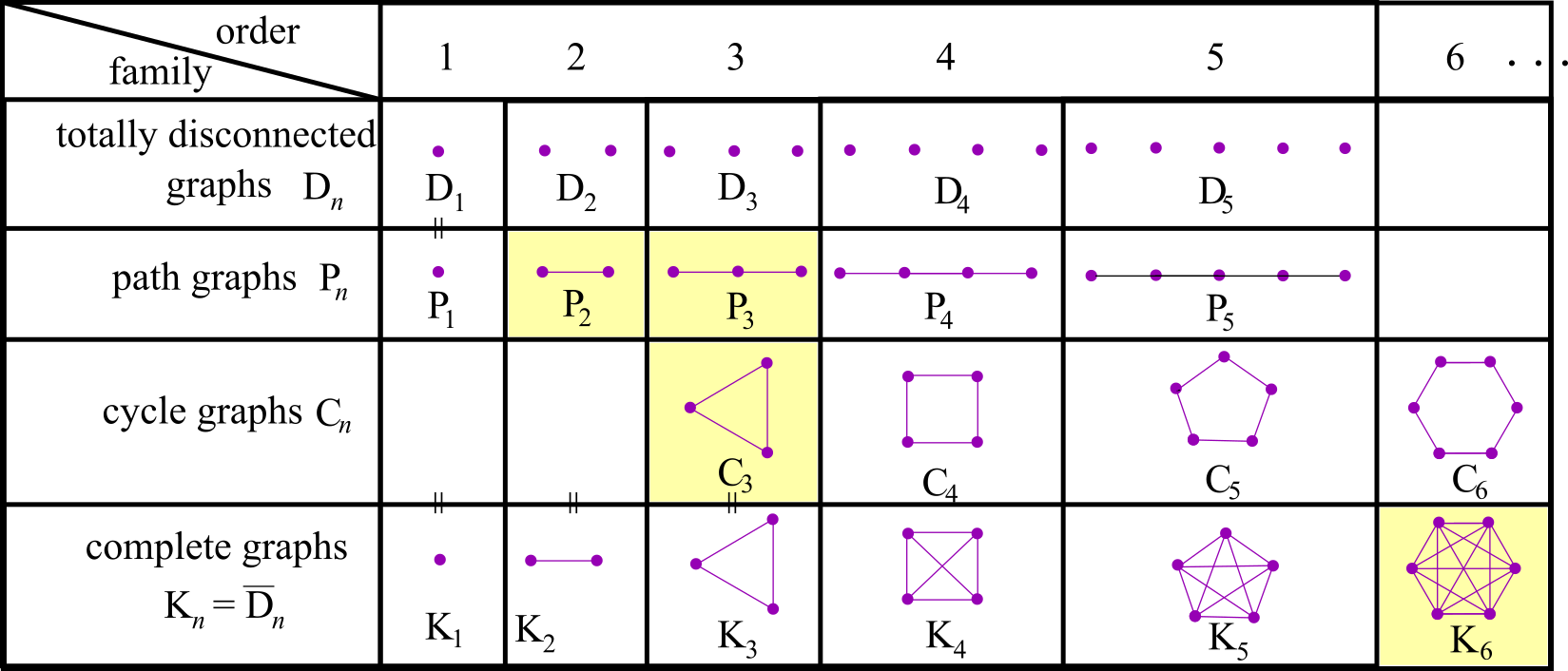}
\caption[Text der im Bilderverzeichnis auftaucht]{        \footnotesize{Totally disconnected, path, cycle and complete graphs. 
Throughout this Appendix's figures, a yellow background indicates examples of graphs which are shape-theoretically realized within this article, 
and an orange background for significant subgraphs thereof as discussed in the current article.  
} }
\label{Graph-1} \end{figure}          }

\m 

\n{\bf Remark 5} The complete graph has the maximal number of edges which a graph of order $|\mG|$ can support, 
\be 
e(\mG)\mbox{-max}  \:=  \frac{|\mG|\{|\mG| - 1\}}{2}                                           \m . 
\ee
\n{\bf Definition 8} A graph's edge saturation is the proportion of the maximal total number of edges realized,  
\be 
Sat   \:=  \frac{e(\mG)}{e(\mG)\mbox{-max}} 
        =  \frac{2 \, e(\mG)}{|\mG|\{|\mG| - 1\}}                                              \m . 
\ee 
\n{\bf Definition 9} A graph H is a {\it subgraph} of a graph G if H's vertex set and edge set are subsets of G's vertex set and edge set respectively.
%

\m 

\n{\bf Definition 10} An {\it edge subdivision} involves removing an edge between 2 vertices v, w, introducing a new vertex u and joining v to u and u to v with new edges. 
A {\it subdivision} alias {\it homeomorph} H of a graph G is a graph obtained from G via a sequence of edge subdivisions. 

\m 

\n{\bf Definition 11} A {\it tree} is a connected graph containing no cycle subgraphs.

\subsection{Graph complements and graph cones}\label{G-2}

\n{\bf Structure 1} The {\it complement} $\overline{\mG}$ of a graph $\mG$  
is a graph which has edges between precisely those vertex pairs which have no edges between them in $\mG$.  

\m 

\n For example, the complete graph $\mK_n$ can also be characterized as the complement of the totally disconnected graph $\mD_n$.  

\m 

\n{\bf Remark 6} For $|\mG|$ the number of vertices of the graph $\mG$, the maximum possible number of edges is 

\m 

\n{\bf Remark 7} The average of a graph and its complement has edge number 
\be 
\langle \me(\mG) \rangle  \es  \frac{1}{2}  \left(  \me(\mG) + \me(\overline{\mG}) \right)   
                          \es  \frac{1}{2}  \left(  \me(\mG) + \frac{|\mG|\{|\mG| - 1\}}{2} - \me(\mG) \right) 
						  \es  \frac{|\mG|\{|\mG| - 1\}}{4} 
                          \es  \frac{\me_{\sm\sa\sx}(\mG)}{2}						                          \m .
						  \label{Semi-Sat}
\ee 
If a graph $\mG$'s saturation is over half of this value, it is usually more straightforward to characterize, recognize and depict its complement $\overline{\mG}$.  

\m 

\n{\bf Structure 2} Graphs can be considered modulo complementation for some purposes.
We denote the equivalence class consisting of $\mG$ and $\overline{\mG}$ by $[\mG]$.

\m

\n{\bf Definition 12} The {\it cone} $\mC(\mG)$ {\it over a graph} $\mG$ is a graph of order 
\be
|\mC(\mG)| = |\mG| + 1
\ee 
whose extra vertex -- the cone vertex -- has edges leading to all of $\mG$'s vertices (so its valency is $|\mG|$).   

\m 

\n{\bf Example 1} 
\be
\mC(\mD_n) = \mS_n                                                                                                      \m : 
\ee 
the $n$-{\it star graph} ($n$-pointed star, of order $n + 1$).  
Fig \ref{Graph-2} shows that the first nontrivial star is $\mS_3$, an alias for which is the {\it claw graph}; this is realized as the (3, 2) topological shape space.  
Note that the stars are the most branched trees of a given order $|\mG|$, to the paths being the least branched trees.  

\m 

\n{\bf Example 2} 
\be
\mC(\mP_n) = \mF_{n - 1}                                                                                                \m : 
\ee
the $(n - 1)$-{\it fan graph} (hand fan with $n - 1$ folds, of order $n + 1$).  
Fig \ref{Graph-2} also shows that the first nontrivial fan is $\mF_2$, an alias for which is the {\it diamond graph}; 
                                               an alias for $\mF_3$                     is the {\it gem graph}; 
												 both of these are realized as topological relational spaces.  

\m 

\n{\bf Example 3} 
\be 
\mC(\mC_n) = \mW_n     \m : 
\ee
the $n$-{\it wheel graph} (wheel with $n$ spokes, of order $n + 1$).  
Fig \ref{Graph-2} also points to the first nontrivial wheel graph being $\mW_4$.

\m 

\n{\bf Example 4} Since 
\be
\mC(\mK_n) = \mK_{n + 1}  \m , 
\ee  
in this case coning does not generate any new graphs.  
%
{            \begin{figure}[!ht]
\centering
\includegraphics[width=0.7\textwidth]{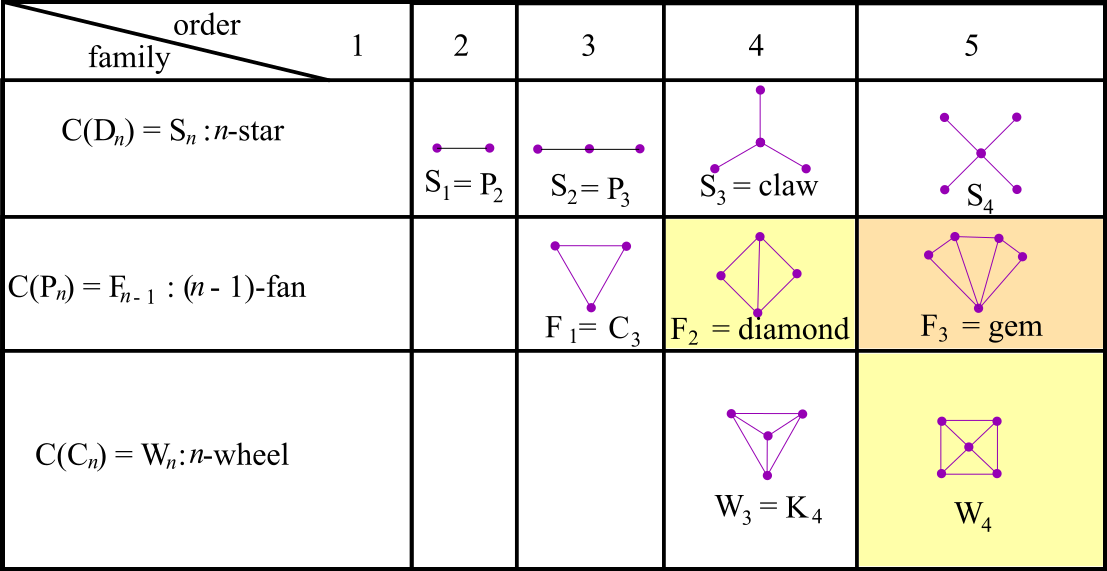}
\caption[Text der im Bilderverzeichnis auftaucht]{        \footnotesize{Cones over totally disconnected, path and cycle graphs: star, fan and wheel graphs.} }
\label{Graph-2} \end{figure}          }

\m 

\n{\bf Lemma 1} Cone graphs admit a trivial characterization in terms of graph complements, 
\be 
\overline{\mC(\mG)} \es \overline{\mG} \, \disjoint \, \mD_1                                                            \m .  
\label{Con}
\ee 

\m 

\n{\bf Definition 13} The {\it cone of a cone graph},  $\mC(\mC(\mG))$, 
is a graph of order 
\be
|\mC(\mC(\mG))| = |\mG| + 2
\ee 
in which two vertices have edges leading to all of $\mG$'s vertices {\sl and} to each other (so these two vertices have valency $|\mG| + 1$).

\m 

\n Recursively,  the k{\it th cone of a graph},  $\mC^k(\mG)$, 
is a graph of order 
\be 
|\mC^k(\mG)| = |\mG| + k
\ee 
in which $k$ vertices have edges leading to all of $\mG$'s vertices {\sl and} to each other (so these $k$ vertices have valency $|\mG| + k - 1$).

\m 

\n{\bf Definition 14} The {\it suspension graph}, $\mS(\mG)$, 
of a given graph $\mG$ has all of $\mG$'s edges and vertices plus two vertices which are each joined by $|\mG|$ further edges, one to each vertex of $\mG$.  
 
\m 
 
\n{\bf Remark 8} These are not the same as cones over cones, since the two suspension points are not themselves joined by an edge, giving the following characterization. 

\m 

\n{\bf Corollary 1} Cone of a cone graphs admit trivial characterization in terms of graph complements:   
\be 
\overline{\mC(\mC(\mG))} \es \overline{\mG} \, \disjoint \,  \mD_1 \, \disjoint \,  \mD_1                                            \m   
\ee 
and
\be 
\overline{\mC^k(\mG)} \es \mG \, \disjoint \, \mD_k  \m = \m  \mG \coprod_{i = 1}^k \mD_1                                            \m . 
\ee 

\m 

\n{\bf Lemma 2} (2-)suspension graphs admit a trivial characterization in terms of graph complements as well:  
\be 
\overline{\mC(\mG)} \es \overline{\mG} \, \disjoint \,  \mP_2                                                                        \m .
\label{Susp} 
\ee 
%
{            \begin{figure}[!ht]
\centering
\includegraphics[width=0.9\textwidth]{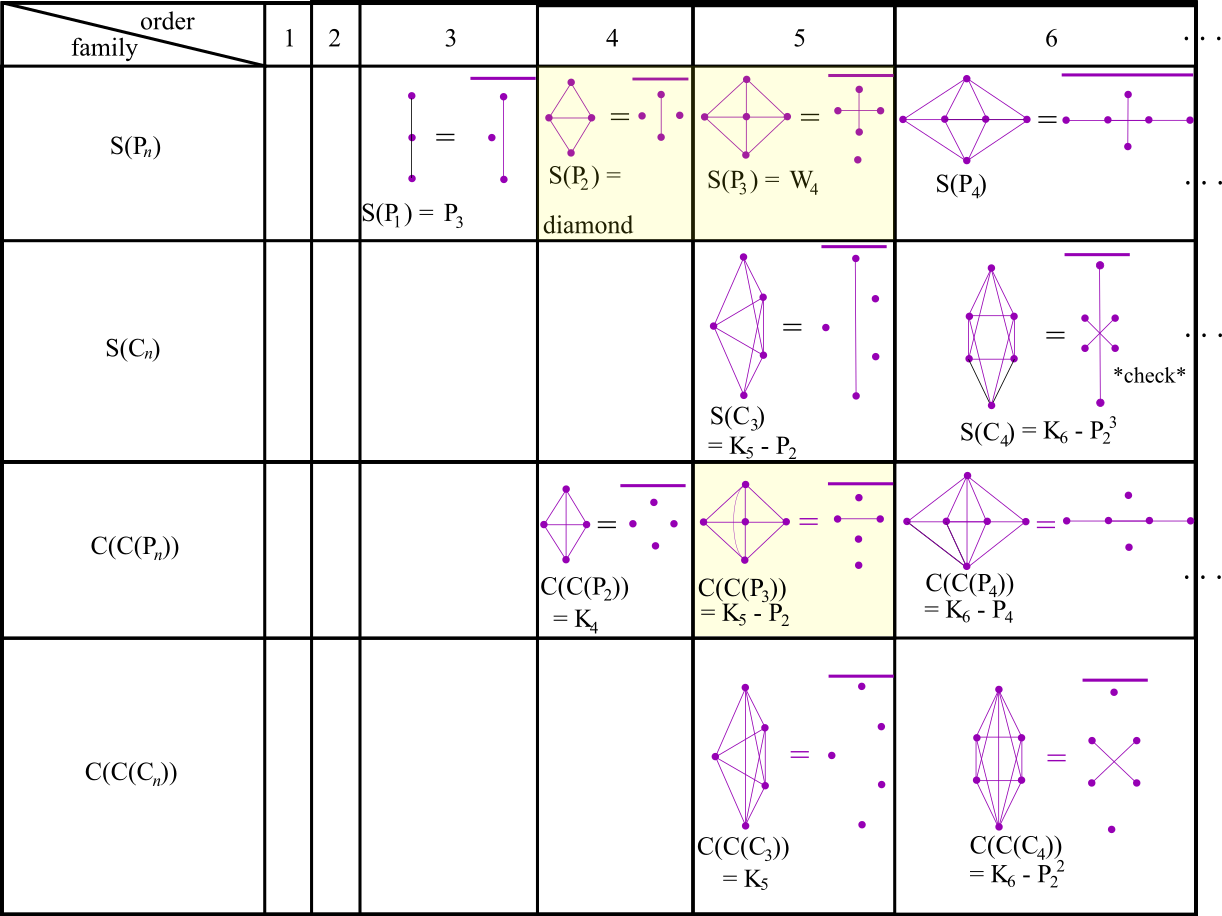}
\caption[Text der im Bilderverzeichnis auftaucht]{        \footnotesize{Cones of cones of graphs, 
including their often more readily identifiable complements.} }
\label{Graph-3} \end{figure}          }


\n{\bf Definition 15} {\it Multipartite graphs} are graphs whose vertices can be partitioned into subsets such that there are no edges within any one given subset.
The simplest nontrivial version of this is for a partition into two subsets: {\it bipartite graphs} (Fig \ref{Graph-4}), the best-known of which are complete within this restriction.  
%
{            \begin{figure}[!ht]
\centering
\includegraphics[width=0.8\textwidth]{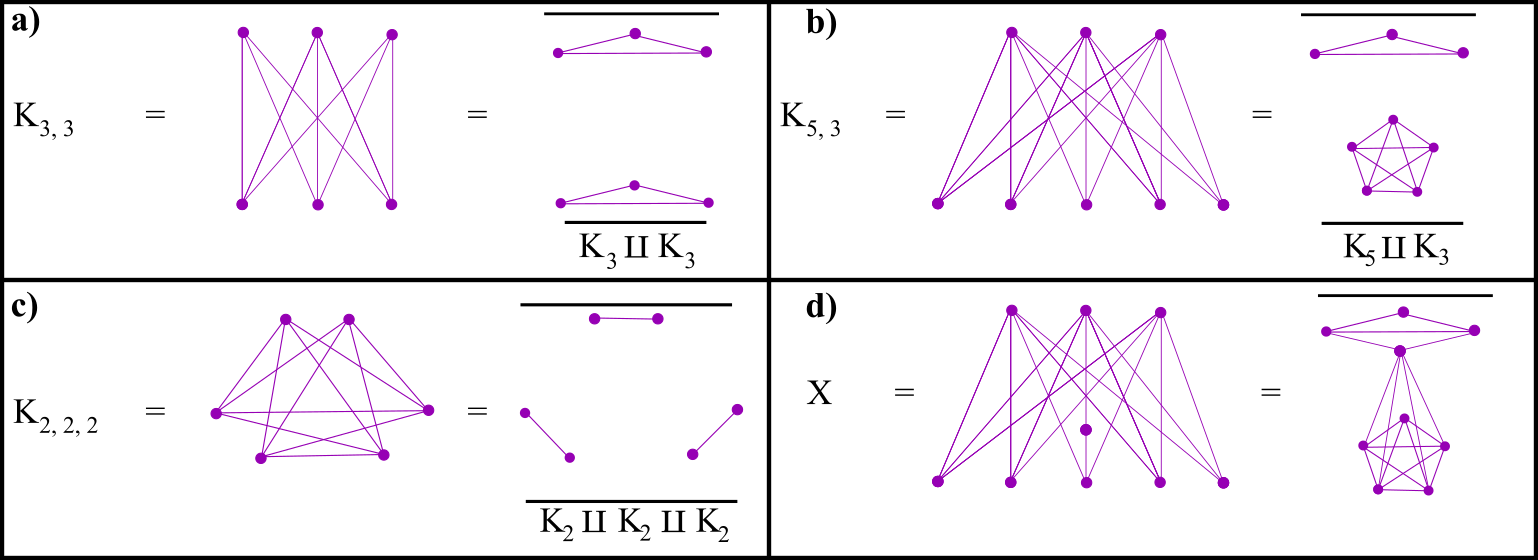}
\caption[Text der im Bilderverzeichnis auftaucht]{        \footnotesize{a), b) Complete bipartite and c) tripartite graphs, in each case alongside their complements.
The homeomorph d) of b) is used in a subsequent proof.
} }
\label{Graph-4} \end{figure}          }

\subsection{Graph-theoretical triviality criteria for Shape-theoretic use}\label{G-3}

\n According to the manner in which we presented the following in Appendix A.1, we have the following criteria.  

\m 

\n{\bf Triviality Criterion 0} The empty graph -- no edges or vertices -- is graph-theoretically trivial. 

\m

\n{\bf Triviality Criterion 1} The disconnected point $\mD_1$ is graph-theoretically trivial. 

\m

\n{\bf Triviality Criterion 2} Paths $\mP_n$ are graph-theoretically trivial. 

\m 

\n{\bf Triviality Criterion 3} Cycles $\mC_n$ are graph-theoretically trivial. 

\m

\n{\bf Triviality Criterion 4} Disjoint unions of trivial graphs are themselves trivial. 
This is because they can be treated component by component, with each component itself requiring less mathematics than Graph Theory to treat. 

\m  

\n{\bf Triviality Criterion 5} A graph $\mH$ is trivial if its complement $\overline{\mH}$ is.  

\m 

\n Applied to Criterion 1), this gives complete graphs $\mK_n = \overline{\mD}_n$ to be trivial as well.

\m 

\n{\bf Triviality Criterion 6} Cones and suspensions of trivial graphs are graph-theoretically trivial. 

\m 

\n{\bf Remark 9} Thus in particular stars, fans and wheels are trivial, by being $\mC(\mD_n)$, $\mC(\mP_n)$ and $\mC(\mC_n)$ respectively. 

\m 

\n{\bf Definition 16} The {\it desuspension} $\mG_{\sD}$ of a graph $\mG$ is the end-product of sequentially removing all suspension points.
The {\it deconing} subcase of this is the one most commonly encountered in the current article.   

\m 

\n{\bf Remark 10} Criteria 5 and 6 bear relation by eqs (\ref{Con}, \ref{Susp}). 
We do not consider $\mP_n$ and $\mC_n$ analogues of this since what Topological Relational Theory produces is cone graphs, double-cone graphs, 2-point suspension graphs...
Coning is relevant to appending scale, and to G as well as O being adjacent to all other vertices in both $d \geq 2$ and $\Top\mbox{-}\Leib$.  
One basic idea is that there are 3 primary types of trivial graphs and that combinations of disjoint unions, 
complements and suspensions thereof maintain a somewhat weaker sense of triviality. 
Many ways in which graphs become interesting do not occur among these graphs.
An exception occurs in the study of planar graphs, to which we now turn.

\subsection{Graph planarity}\label{G-4}

\n{\bf Definition 17} A graph is {\it planar} \cite{Graphs-2} if it can be embedded in a plane. 
This means that it can be drawn on a piece of paper such that its edges intersect at their vertex end-points alone.  

\m 

\n{\bf Definition 18} A {\it Kuratowski graph} is any subdivision of $\mK_5$ or $\mK_{3,3}$. 
A graph possesses a $\mK$-subgraph if it contains a Kuratowski subgraph as a subgraph.  

\m 

\n{\bf Theorem 1 (Kuratowski's subgraph criterion for planarity)} \cite{Graphs-1} A finite graph is planar iff it does not contain a subgraph that subdivides $\mK_5$ or $\mK_{3, 3}$.   

\m 

\n The following simple diagnostic readily follows from the preceding and Fig \ref{Graph-4}.a)-b). 

\m

\n{\bf Remark 11} $\mK_5$ and $\mK_{3,3}$ are moreover a sequential cone and a suspension, and also the complements of $\mD_5$ and $\mK_3 \, \disjoint \, \mK_3$, 
in which ways they are trivial according to previous criteria. 
These complements moreover underline the following simple Corollary.  

\m 

\n{\bf Corollary 2} If a graph's complement contains a $\mD_5$ subgraph or a $\mK_3 \, \disjoint \, \mK_3$ subgraph, in each case with no adjacent edges joining its components, 
then the graph is planar. 

\m

\n{\bf Remark 12} It should however be noted that the above triviality criteria for $\mK_5$ and $\mK_{3,3}$ themselves need not be inherited under the 
`is present as a subgraph' or homeomorph generalizations.
The case of homeomorphs is not covered by the Corollary either, by which it is a weak Corollary albeit a sufficient one for the current Article's purposes. 

\m 

\n Various further (non)triviality criteria are then as follows.  

\m 

\n{\bf Triviality Criterion 7.a)} The graph is planar. 

\m 

\n{\bf Triviality Criterion 7.b)} The [graph] is [planar] meaning that both it and its complement are planar. 
Nontriviality here is that the [graph] is [nonplanar], meaning that at least one of it and its complement are nonplanar.  

\m 

\n{\bf Non-Triviality Criterion 7.c)} Both the graph and its complement are nonplanar.   

\m

\n{\bf Triviality Criterion 8} This is the version of Criterion 7 after desuspending,  of shape-theoretic interest as applied to $\Top-\Leib$'s residue graph.  
This includes in the restricted sense of targetting particular privileged-label suspensions only, i.e.\ specifically removing some subset of O, 2 and G and points.   
One use of this is having a criterion of nontriviality that is the same for both memebers of a Jacobi pair of relational theories. 
Another use is in considering the collision structure, 
for which G's are excised (whether or not O's are excised as well to pass to a normalizable shape theory within a Jacobi pair of relational theories).

\subsection{Graph-Theoretic order bounds on the triviality criteria}\label{G-5}

\n{\bf Lemma 3}. 

\m 

\n Criterion 0) requires $|\mG| \geq 1$. 

\m 

\n Criterion 1) requires $|\mG| \geq 2$. 

\m 

\n Criterion 2) requires $|\mG| \geq 3$. 

\m 

\n Criteria 3) and 4) require $|\mG| \geq 4$.  

\m 

\n Criteria 5), 6) and 7.a) require $|\mG| \geq 5$.  

\m

\n Criterion 7.b) requires $|\mG| \geq 5$ to occur, and becomes obligatory for $|\mG| \geq 9$. 

\m 

\n Criterion 7.c) requires $|\mG| \geq 8$ to occur, but never becomes obligatory.

\m

\n Criteria 8.a)-c)) require $|\mG_{\sD}| \geq$ 6, 6 and 9 and 9 respectively.   

\m 

\n\underline{Proof}. 
0) to 3) are just identifying the first point, path, cycle, and finally any of $\mK_4$, diamond, co-diamond, paw, co-paw, claw, co-claw and co-square (Fig \ref{Graph-6}.a). 

\m 

\n For 4), $\mK_4$, diamond, paw and claw continue to suffice.   

\m 

\n 5) however overcomes all of 3)'s examples since co-square, co-paw, co-claw and co-diamond are all disjoint unions of trivial graphs, as is $\mD_4 = \overline{\mK_4}$.  
Yet $|\mG|$ = 5 suffices to have a choice of two counterexamples: the `chair, co-chair' and `banner, co-banner' pairs (Fig \ref{Graph-6}.b).  
Both of these additionally serve for 6) since neither is a coning nor a suspension.

\m

\n 7.a) Nonplanarity is only possible if a $\mK_5$ or $\mK_{3,3}$ subgraph is present, by which $|\mG| \geq 5$ is clear; this also gives the first part of 7.b). 
For 8.a) and the first part of 8.b), $\mK_5$ is however a sequential coning, though its unique $|\mG| = 6$ homeomorph is not, securing $|\mG| \geq 6$.  

\m 

\n 7.c) Nonplanarity--and--non-co-planarity first occurs for the complementary pair $\mK_{5,3}$ and $\mK_5 \, \disjoint \,  \mK_3$ (Fig (\ref{Graph-4}.c) 
as the unique $|\mG| = 8$ example.
The first graph in this pair is however a suspension (3-point suspension over $\mD_5$).
It moreover also only takes a 1-vertex homeomorph to break this (Fig \ref{Graph-4}.d), securing $|\mG| = 9$ as minimal to have a graph--complement pair 
neither of which are planar or suspensions: 8.c).    

\m 

\n For the second part of 7.b), planar graphs are forced to have nonplanar complements for $|\mG| = 9$ \cite{Battle}.
(This also secures the second part of 8.b), since some of the [graphs] in question are not [suspensions].)   
While this $|\mG| = 9$ result is not straightforward to prove, the $N \geq 11$ bound follows trivially from Euler's formula \cite{Graphs-1}.  
$\Box$ 
%
{            \begin{figure}[!ht]
\centering
\includegraphics[width=0.8\textwidth]{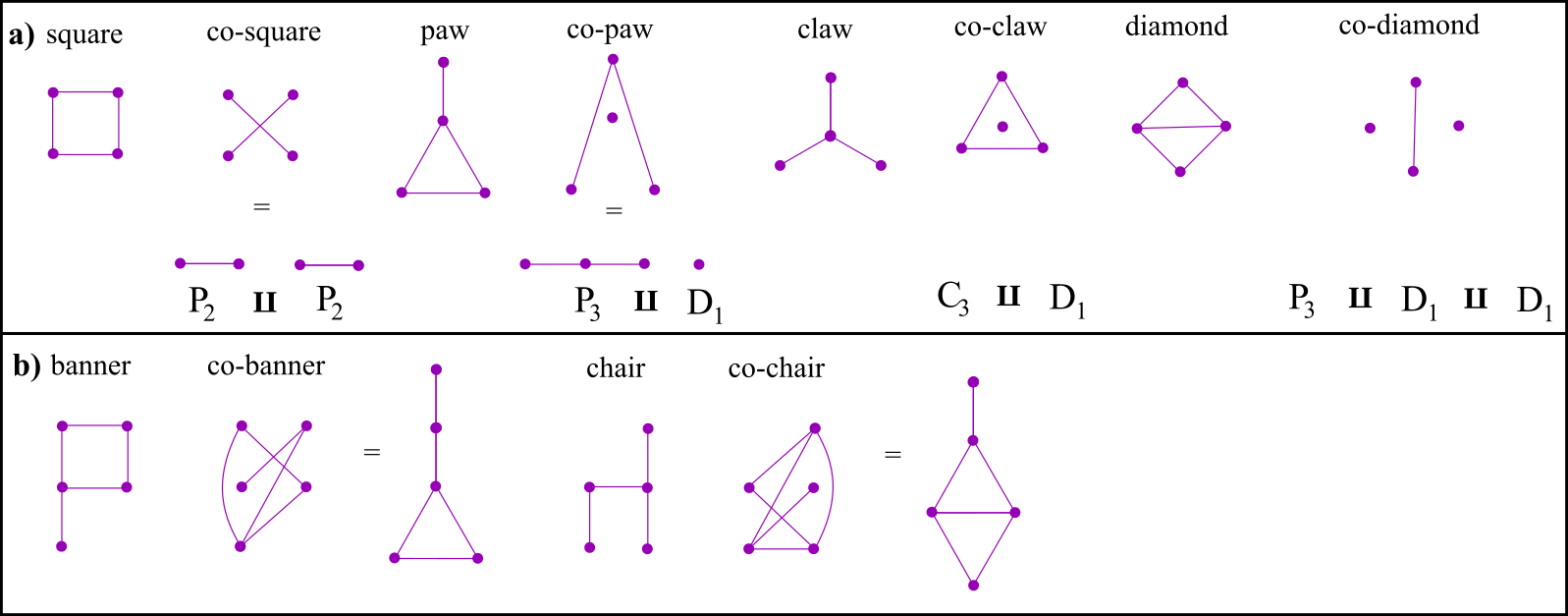}
\caption[Text der im Bilderverzeichnis auftaucht]{        \footnotesize{a) square, co-square, claw, co-claw, paw and co-paw.

\m

\n b) Chair, co-chair, banner and co-banner.

\m 

\n c) $\mK_{5,3}$ and its complement $\mK_5 \, \disjoint \,  \mK_3$.  

\m 

\n d) A minimally [nonplanar] graph that is not a suspension either. 
} }
\label{Graph-6} \end{figure}          }

\vspace{10in}

\section{Supporting Lattice Theory}\label{App-L}

\n{\bf Structure 1} A {\it binary relation} $R$ on a set $\FrX$ is a property that each pair of elements of $\FrX$ may or may not possess.  
We use $a \, R \, b$ to denote `$a$ and $b$ $\in \FrX$ are related by $R$'. 

\mbox{ }

\n{\bf Structure 2} Some basic properties that a binary relation $R$ on $\FrX$ might possess are as follows (i$\forall \, a, b, c \in \FrX$ wherever applicable).  

\m 

\n i) {\it Reflexivity}:  $a \, R \, a$. 

\m 

\n ii) {\it Antisymmetry}: $a \, R \, b$ and $b \, R \, a \Rightarrow a = b$.      

\m

\n iii) {\it Transitivity}: $a \, R \, b$ and $b \, R \, c \Rightarrow a \, R \, c$.  

\m 

\n iv) {\it Totality}: that one or both of $a \, R \, b$ or $b \, R \, a$ holds, i.e.\ all pairs are related.  

\mbox{ }  

\n{\bf Definition 1} A binary relation $R$ is a {\it partial ordering}, which we denote by $\preceq$, if $R$ is reflexive, antisymmetric and transitive. 
$R$ is moreover a {\it total ordering} alias {\it chain} if it is both a partial order and total.  

\m

\n Example 0) $\leq$ acting on the real numbers is a total ordering, whereas $\subseteq$ acting on sets as `is a subset of' is a partial ordering. 

\m

\n{\bf Definition 2} A {\it poset} is a set equipped with a partial order, $\langle \FrX, \preceq \rangle$.  

\m 

\n{\bf Definition 3} An {\it antichain} is a subset $\FrA \subset \FrX$ such that no two elements of $\FrA$ are related.   

\m

\n{\bf Remark 1} (Small finite) posets are conveniently represented by Hasse diagrams \cite{Pure-Lattice}; see Fig \ref{Part-Latt} for some examples of these.

\m

\n{\bf Definition 4} A {\it lattice} $\lattice$ \cite{Pure-Lattice, StanleyBook} 
is a poset for which each pair of elements possesses a {\it join} $\lor$ (least upper bound) and a {\it meet} greatest lower bound $\land$.

\m 

\n The study of orders constitutes {\it Order Theory} \cite{Pure-Lattice, StanleyBook},  
with {\it Lattice Theory} the itself-rich study of the specialization of this to lattices.

\m 

\n{\bf Definition 5} The {\it dual} of a given lattice is another lattice in which $\lor$ and $\land$'s statuses are reversed.  
The corresponding Hasse diagrams are upside-down relative to each other. 
If the arrows are furthermore reversed, the so-called {\it antitone dual} is realized.  

\m 

\n{\bf Definition 6} An element 1 of $\lattice$ is a {\it top} alias {\it unit element}         if 
\be 
\forall \, l \,  \in \,  \lattice   \mma 
l \preceq 1                         \m .
\ee 
An element 0 of $\lattice$ is a {\it bottom}, {\it null} or {\it zero element} if 
\be 
\forall \, l \,  \in \,  \lattice   \mma 
0 \preceq l                         \m . 
\ee
A lattice possessing both of these is termed a {\it bounded lattice}.

\m 

\n{\bf Definition 7} A {\it lattice morphism} is an order-, join- and meet-preserving bijection between lattices.  

\m

\n{\bf Remark 2} One can moreover view lattices as a specialization of directed graphs, now with some lattice-theoretic restrictions on both which shapes of graph are acceptable 
and on how these are to be embedded so that the corresponding graph is a valid Hasse diagram.  

\m 

\n{\bf Example 1} Partition refinement for a fixed $N$ constitutes a boundedlattice.
Its top element is all the objects in a single box, whereas its bottom element is its elements split one per box into $N$ boxes. 
Fig \ref{Part-Latt} depicts these lattices for the $N = 1$ to 8 cases required by the current article.  
%
{            \begin{figure}[!ht]
\centering
\includegraphics[width=1.0\textwidth]{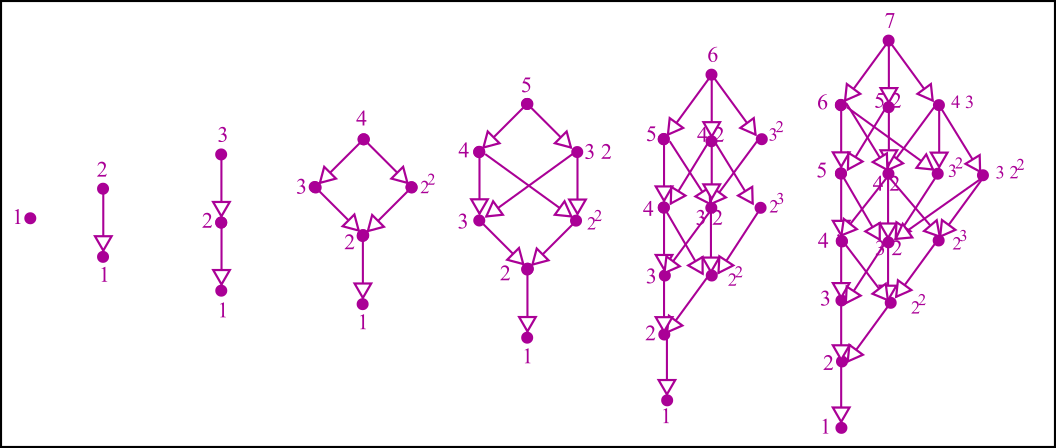}
\caption[Text der im Bilderverzeichnis auftaucht]{        \footnotesize{Partition refinement lattices on 1 to 7 points.} }
\label{Part-Latt} \end{figure}          }

\m 

\n{\bf Example 2} The set of subsets of a fixed finite set $\FrX$ forms a bounded lattice $\lattice(\FrX)$ under the ordering `is a subset of'.   
The top and bottom elements here are $\FrX$ and $\emptyset$, the meet is intersection and the join is the smallest subspace containing a pair of spaces. 

\m 

\n We next consider expanding Example 2 to sets with further structure, in particular from finite sets to finite groups as follows. 

\m 

\n{\bf Example 3} The subgroups of a group form a bounded lattice $\lattice(\lFrg)$ under the ordering `is a subgroup of'.  
The top and bottom elements here are the whole group $\lFrg$ and the trivial group $id$, and the join is the subgroup generated by their union.

\m 

\n{\bf Example 4} Given an object space $\FrO\mb$ of objects $\FrO$ that a group $\lFrg$ acts upon, some of the subgroups of $\lFrg$ may act identically. 
In this case, a smaller bounded lattice can be formed, of {\it distinct subgroup actions} on $\FrO\mb$, 
\be 
\lattice(\s{\rightarrow}{\lFrg} \FrO\mb) \m .
\ee    
Its top and bottom elements are the whole group's action on $\FrO\mb$ and the trivial action on $\FrO\mb$ (i.e.\ $id$ acting on each $\FrO$ to simply return that $\FrO$ again).  

\m 

\n{\bf Example 5} A subexample of the previous is for $\lFrg$ acting on $\FrO\mb$ to produce a quotient, 
by which the quotients under distinctly-acting subgroups themselves form a bounded lattice, 
\be 
\lattice 
\left(
\frac{\FrO\mb}{\FrH}, {\FrH} \in \lattice(\s{\rightarrow}{\lFrg} \FrO\mb)
\right) \m . 
\ee  
The top element here is $Ob$ itself: the least quotiented space, whereas the bottom element is 
\be
\frac{\FrO\mb}{\lFrg}  \m : 
\ee 
the most quotiented space. 
[For configuration spaces or phase spaces, one often says `reduced' rather than `quotiented'.]
This lattice, as presented, is the antitone dual of $\lattice(\s{\rightarrow}{\lFrg} \FrO\mb)$.  

\m 

\n In particular, in the case of quotienting a relational space by discrete groups $\Gamma$, 
this accords lattice-theoretic significance to Leibniz spaces (metric-level or topological-level).  

\vspace{10in}

\end{appendices}

\vspace{10in}


\end{document}